\documentclass[reprint, groupedaddress, showpacs, amsmath, amssymb,aps,prd,nofootinbib]{revtex4-1}
\usepackage[bottom]{footmisc} 
\usepackage[utf8]{inputenc}
\usepackage{footnote}
\usepackage{empheq}
\usepackage{amsmath}
\usepackage{mathtools}
\usepackage{amsfonts}
\usepackage{amssymb}
\usepackage{IEEEtrantools}
\usepackage{graphicx}
\usepackage{float}
\usepackage{hyperref}
\usepackage[framemethod=tikz]{mdframed}
\usepackage{lipsum}
\usepackage{tikz-feynman}
\usepackage{slashed}
\usepackage{comment}
\usepackage{xcolor}
\usepackage{natbib}
\usepackage{url}
\usepackage{textcase}
\usepackage{bm}
\usepackage{amsthm}
\usepackage{verbatim}
\usepackage{epsfig}
\usepackage{slashed}

\tikzfeynmanset{compat=1.1.0}
\newcommand{\p}[4]{\left(#1\right)^{[#2,#3]}_{+(#4)}}
\newmuskip\pFqmuskip
\newcommand\msb{\overline{\text{MS}}}
\newcommand*\pFq[6][8]{%
  \begingroup 
  \pFqmuskip=#1mu\relax
  \mathcode`\,=\string"8000
  \begingroup\lccode`\~=`\,
  \lowercase{\endgroup\let~}\pFqcomma
  {}_{#2}F_{#3}{\left[\genfrac..{0pt}{}{#4}{#5};#6\right]}%
  \endgroup
}

\newcommand{\pFqcomma}{\mskip\pFqmuskip}
\newcommand{\Blue}[1]{\textcolor{black}{#1}}
\newcommand{\Red}[1]{\textcolor{black}{#1}}

\newcommand{\blue}[1]{\textcolor{black}{#1}}

\newcommand{\blueX}[1]{\textcolor{black}{#1}}
\newcommand{\BlueX}[1]{\textcolor{black}{#1}}

\makeatletter
\newsavebox\myboxA
\newsavebox\myboxB
\newlength\mylenA

\newcommand*\xoverline[2][0.75]{%
    \sbox{\myboxA}{$\m@th#2$}%
    \setbox\myboxB\null
    \ht\myboxB=\ht\myboxA%
    \dp\myboxB=\dp\myboxA%
    \wd\myboxB=#1\wd\myboxA
    \sbox\myboxB{$\m@th\overline{\copy\myboxB}$}
    \setlength\mylenA{\the\wd\myboxA}
    \addtolength\mylenA{-\the\wd\myboxB}%
    \ifdim\wd\myboxB<\wd\myboxA%
       \rlap{\hskip 0.5\mylenA\usebox\myboxB}{\usebox\myboxA}%
    \else
        \hskip -0.5\mylenA\rlap{\usebox\myboxA}{\hskip 0.5\mylenA\usebox\myboxB}%
    \fi}
\makeatother

\begin{document}
\title{One-loop Matching Factors for Singlet Quasi-Parton Distribution Functions in the Hybrid-Ratio Scheme}
\author{Yi-Xian Chen}
\email{w05217@gmail.com}
\affiliation{Department of Physics and Center for Theoretical Sciences,
National Taiwan University, Taipei, Taiwan 106}
\affiliation{Physics Division, National Center for Theoretical \Blue{Physics}, Taipei 10617, Taiwan}
\author{Jiunn-Wei Chen}
\email{jwc@phys.ntu.edu.tw}
\affiliation{Department of Physics, Center for Theoretical \Blue{Physics},
and Leung Center for Cosmology and Particle Astrophysics,
National Taiwan University, Taipei, Taiwan 106}
\affiliation{Physics Division, National Center for Theoretical Sciences, Taipei 10617, Taiwan}
\affiliation{Center for Gravitational Physics and Quantum Information,
Yukawa Institute for Theoretical Physics,
Kyoto University, Kyoto 606-8502, Japan}

\begin{abstract}
The one loop matching kernels between parton distribution functions (PDFs) for parton $i=u,d,s,g$ and their corresponding quasi-PDFs 
are computed at one loop in the hybrid-ratio scheme. We found that, in addition to the conservation of the quasi-quark number for each flavor, the second moment $\langle x \rangle_{\tilde{i}}=\langle x \rangle_i$ of quasi-PDF of \Blue{parton $i$ (denoted as $\tilde{i}$)} and PDF of parton $i$ is the same in our approach. This is demonstrated numerically using the CTEQ14 global analysis as input. 
\end{abstract}
\maketitle
\section{Introduction}

Parton distribution functions (PDFs) are fundamental structures of hadrons \Blue{in terms of} quarks and gluons distributions. \Blue{Many mid-energy facilities worldwide} are trying to determine these structures and their three-dimensional generalizations, such as a future electron-ion collider (EIC) \cite{AbdulKhalek:2022erw},
at Brookhaven. EIC is a multi-billion US-dollar facility scheduled to take data in about 10 years. The previous determination of proton PDFs is already applied to physics searches beyond the Standard Model in energy-frontier experiments \Blue{such as} LHC.

Large-momentum effective theory (LaMET)~\cite{Ji:2013dva,Ji:2014gla} \Blue{enables} computation of the Bjorken-$x$ dependence of hadron PDFs on a Euclidean lattice. It is complementary to experiments, especially in kinematic regions that are difficult to access in experiments.
LaMET relates equal-time spatial correlators, whose Fourier transforms are called quasi-PDFs, to PDFs in the limit of infinite hadron momentum.
For large but finite momenta accessible on a realistic lattice, LaMET relates quasi-PDFs to physical ones through a factorization theorem \Blue{(see Ref. \cite{Ma:2017pxb,Izubuchi:2018srq,Liu:2019urm} for the proofs)}.

Since the proposal of LaMET, significant advancements have been achieved in the theoretical understanding of the formalism.~\cite{Xiong:2013bka,Ji:2015jwa,Ji:2015qla,Xiong:2015nua,Ji:2017rah,Monahan:2017hpu,Stewart:2017tvs,Constantinou:2017sej,Green:2017xeu,Izubuchi:2018srq,Xiong:2017jtn,Wang:2017qyg,Wang:2017eel,Xu:2018mpf,Chen:2016utp,Zhang:2017bzy,Ishikawa:2016znu,Chen:2016fxx,Ji:2017oey,Ishikawa:2017faj,Chen:2017mzz,Alexandrou:2017huk,Constantinou:2017sej,Green:2017xeu,Chen:2017mzz,Chen:2017mie,Lin:2017ani,Chen:2017lnm,Li:2016amo,Monahan:2016bvm,Radyushkin:2016hsy,Rossi:2017muf,Carlson:2017gpk,Ji:2017rah,Briceno:2018lfj,Hobbs:2017xtq,Jia:2017uul,Xu:2018eii,Jia:2018qee,Spanoudes:2018zya,Rossi:2018zkn,Liu:2018uuj,Ji:2018waw,Bhattacharya:2018zxi,Radyushkin:2018nbf,Zhang:2018diq,Li:2018tpe,Braun:2018brg,Detmold:2019ghl,Sufian:2020vzb,Shugert:2020tgq,Green:2020xco,Braun:2020ymy,Lin:2020ijm,Bhat:2020ktg,Chen:2020arf,Ji:2020baz,Chen:2020iqi,Chen:2020ody,Alexandrou:2020tqq,Fan:2020nzz,Ji:2020brr,Chen:2018xof,Lin:2018qky,Liu:2018hxv,Liu:2020rqi}.
The method has been utilized in lattice calculations of PDFs to analyze the up and down quark content of the nucleon~\cite{Lin:2014zya,Chen:2016utp,Lin:2017ani,Alexandrou:2015rja,Alexandrou:2016jqi,Alexandrou:2017huk,Chen:2017mzz,Lin:2018pvv,Alexandrou:2018pbm,Chen:2018xof,Alexandrou:2018eet,Lin:2018qky,Liu:2018hxv,Wang:2019tgg,Lin:2019ocg,Liu:2020okp,Lin:2019ocg,Zhang:2019qiq,Alexandrou:2020qtt},
$\pi$~\cite{Chen:2018fwa,Izubuchi:2019lyk,Lin:2020ssv,Gao:2020ito,Gao:2021dbh} and $K$~\cite{Lin:2020ssv} mesons,
and the $\Delta^+$~\cite{Chai:2020nxw} baryon.
Despite limited volumes \cite{Lin:2019ocg,Liu:2020krc} and relatively coarse lattice spacings, previous state-of-the-art nucleon isovector quark PDFs, derived from lattice data at the physical pion mass~\cite{Lin:2018pvv,Alexandrou:2018pbm,LatticeParton:2022xsd} and the physical-continuum limit (with continuum extrapolations at the physical pion mass)~\cite{Lin:2020fsj}, have demonstrated reasonable agreement with phenomenological results obtained from experimental data.
Building on this success, LaMET has been extended to include twist-three PDFs~\cite{Bhattacharya:2020cen,Bhattacharya:2020xlt,Bhattacharya:2020jfj} and generalized parton distributions (GPDs)~\cite{Bhattacharya:2021oyr}, as well as gluon~\cite{Fan:2018dxu,Fan:2020cpa,Salas-Chavira:2021wui}, strange, and charm distributions~\cite{Zhang:2020dkn}. It has also been applied to meson distribution amplitudes (DAs)~\cite{Zhang:2017bzy,Chen:2017gck,Zhang:2020gaj,Hua:2020gnw,Hua:2022kcm} and GPDs~\cite{Chen:2019lcm,Alexandrou:2020zbe,Lin:2020rxa,Alexandrou:2019lfo,Lin:2021brq,Alexandrou:2021bbo}.\par
Efforts have also been undertaken to generalize LaMET to transverse momentum dependent (TMD) PDFs~\cite{Ji:2014hxa,Ji:2018hvs,Ebert:2018gzl,Ebert:2019okf,Ebert:2019tvc,Ji:2019sxk,Ji:2019ewn,Ebert:2020gxr,Ebert:2022fmh}, to calculate the nonperturbative Collins-Soper evolution kernel~\cite{Ebert:2018gzl,Shanahan:2019zcq,Shanahan:2020zxr,Chu:2022mxh,Shanahan:2021tst}, and to compute soft functions~\cite{Zhang:2020dbb,Li:2021wvl} on the lattice. Additionally, LaMET has \Blue{renewed} interests in earlier approaches~\cite{Liu:1993cv,Detmold:2005gg,Braun:2007wv,Bali:2017gfr,Bali:2018spj,Detmold:2018kwu,Detmold:2021qln,Liang:2019frk} and inspired new ones~\cite{Ma:2014jla,Ma:2014jga,Chambers:2017dov,Radyushkin:2017cyf,Orginos:2017kos,Radyushkin:2017lvu,Radyushkin:2018cvn,Zhang:2018ggy,Karpie:2018zaz,Joo:2019jct,Radyushkin:2019owq,Joo:2019bzr,Balitsky:2019krf,Radyushkin:2019mye,Joo:2020spy,Can:2020sxc,HadStruc:2021wmh}.
For recent reviews on these topics, see~\cite{Lin:2017snn,Cichy:2018mum,Zhao:2020vll,Ji:2020ect,Ji:2020byp,Constantinou:2022yye}.

While the computations of the non-singlet one loop kernels
are more straightforward \cite{Stewart:2017tvs, Chou:2022drv}, there is a lot of subtlety in the singlet case.
The ratio scheme computation of Ref.\cite{Balitsky:2019krf} first concentrated on the quasi-gluon case and started with the Yang-Mills theory with $n_f=0$. In this limit, gluon momentum conservation is guaranteed by the total energy-momentum tensor conservation. In the case of QCD \Blue{with $n_f\ne 0$}, the gluon wave function renormalization has an IR singularity proportional to $n_f$ (see our Eq.(\ref{WF})), which is missing in \cite{Balitsky:2019krf}. 

In Ref.\cite{Wang:2017qyg}, the matching between singlet  PDFs in $\msb$ and quasi-PDFs also in $\msb$ was presented. They agreed with the forward limit of quasi GPD and GPD in \Red{Ref.~\cite{Ma:2022ggj,Ma:2022gty}} done in the same scheme. However, the $\msb$ to $\msb$ matching suffers from the problems described after Eq.(\ref{MC}) in this work.

In Ref.~\cite{Wang:2019tgg}, RI/MOM scheme is used to renormalize the quasi-PDFs. Their $\int dx\ x\tilde{g}(x)$ moment diverges logarithmically and the off-shellness of gluons in the counterterms raises the concern of breaking the gauge symmetry.

Finally, Ref.~\cite{Yao:2022vtp} calculated the one loop matching kernels of hybrid-ratio GPDs to GPDs in $\msb$. Their gluon kernel agreed with that of Ref.\cite{Balitsky:2019krf} in the forward limit in the coordinate space, \Blue{which missed the $n_f$ dependence as explained} above.  

These problems are all addressed in this work with explicit step-by-step derivations. We found that, in addition to the quasi-quark number conservation for each flavor, the second moment $\langle x \rangle_{\tilde{i}}=\langle x \rangle_i$ of quasi-PDF 
parton \Blue{ $i$  (denoted as $\tilde{i}$) }
and PDF of parton $i$ is the same in our approach. This is demonstrated numerically using the CTEQ14 global analysis as input. 


\begin{widetext}

\Blue{\section{General formulation}}

\subsection{Renormalization of singlet quasi-PDFs}

PDFs of hadrons are defined as the Fourier transform of the hadron matrix elements of operators defined on a lightcone. The quark unpolarized PDF is defined as
\begin{equation}
    q(x)\equiv\int_{-\infty}^{\infty}\frac{d\xi^{-}}{4\pi}e^{-ixP^{+}\xi^{-}}\langle P|\bar{\psi}_q(\xi^{-})\gamma^{+}U(\xi^{-},0)\psi_q(0)|P \rangle ,
\end{equation}
where $\psi_q$ is the fermion field operator of flavor $q$ and
the light cone coordinate $\xi^{\pm}=(t\pm z)/\sqrt{2}$.  $P^{\mu}=(P^0,0,0,P^z)$ is the hadron momentum, $x$ is 
the momentum fraction of a parton relative to the hadron momentum. The Wilson line 
\begin{equation}
\label{Wilson}
    U(\xi^{-},0)=\mathcal{P}\ \mathrm{exp}\biggl(-ig \int_{0}^{\xi^{-}}d\Theta^{-} A^{+}(\Theta^{-}) \biggr).
\end{equation}
is a line integral along a path on a lightcone with $\mathcal{P}$ indicating the path ordering of the operator on its right. $A^+\equiv A^+_at^a$ with $t^a$ the generator in the fundamental representation of the color $SU(3)$. 

The gluon unpolarized PDF is defined as
\begin{equation}
    xg(x) \equiv\int_{-\infty}^{\infty}\frac{d\xi^{-}}{\Blue{4}\pi P^+}e^{-ixP^{+}\xi^{-}} 
   \langle P|F^{+i}(\xi^{-})\tilde{U}(\xi^{-},0)\Blue{F^{\ +}_i}(0)|P \rangle
\end{equation}
where \Blue{$F^{\mu\nu}=F^{\mu\nu}_a t^a$}, $F^{\mu\nu}_a=\partial^{\mu}A^{\nu}_a-\partial^{\nu}A^{\mu}_a-gf_{abc}A^{\mu}_bA^{\nu}_c$ is the gluon field strength operator, and $i=x,y$ are the transverse coordinates. The Wilson line $\tilde{U}$ is analogous to the one in Eq.(\ref{Wilson}) except the gauge field is under the adjoint representation.

The quark quasi-PDF operator is defined as
\begin{align}
\label{Oq}
O_{q}(z)=
-i\frac{d}{dz}\bar{\psi}_{q} (z)\gamma^z U(z,0)\psi_{q}(0), 
\end{align}
where $q=\sum_{j=1}^{n_f} q_j/n_f$ is a flavor averaged quark
with $n_f$ the number of light quark flavors. In this fermionic operator,   
$\gamma^t$ instead of $\gamma^z$ is typically used  to avoid mixing with another operator on the lattice due to the breaking of chiral symmetry in the lattice action \cite{Constantinou:2017sej,Chen:2017mie}. We will address this problem in the next publication.
The operator of Eq.(\ref{Oq}) has one more derivative than the non-singlet one proposed in Ref. \cite{Ji:2013dva} such that it has the same mass dimension as the gluon quasi-PDF operator that we will introduce below. The operator is of equal time with the Wilson line going along the $z$ direction:
\begin{equation}
    U(z,0)=\mathcal{P}\ \mathrm{exp}\biggl(-ig \int_{0}^{z'}dz' A^z(z') \biggr).
\end{equation}

The gluon quasi-PDF operator is 
\begin{equation}
\label{O_g}
    O_{g}(z)=F^{z\alpha}(z)\tilde{U}(z,0) F^{\ z}_{\alpha}(0) ,
\end{equation}
where the repeated Lorentz index $\alpha=0,1,2,3$ is summed over.
Their matrix elements of a hadron state with momentum $P^{\mu}=(P^0,0,0,P^z)$ yield the dimensionless equal time correlators \blue{near the $P^z \to \infty$ limit:}
\begin{align}
\label{h-functionX}
h_{i}(\zeta,z^2)=\frac{1}{2 (P^z)^2}\langle P|O_i(z)|P\rangle ,
\end{align}
where $i=g,q$ and $\zeta=z P^z$. Then the quasi-PDFs are Fourier transforms of these correlators
\begin{equation}
\label{Fourier}
   \Blue{x} \tilde{f}_i(x,P^z)\equiv \int_{-\infty}^{\infty}\frac{d\zeta}{2\pi} e^{ix\zeta}h_{i}(\zeta,\frac{\zeta^2}{(P^z)^2}) .
\end{equation}
Note that the negative $x$ part is related the anti-parton ($\bar{i}$) (quasi-)PDF through $\tilde{f}_i(-|x|)=-\tilde{f}_{\bar{i}}(|x|)$ and $f_i(-|x|)=-f_{\bar{i}}(|x|)$.

\blueX{If a lattice regulator is used, near the continuum limit and for $z \ne 0$, we have
\begin{align}
\label{renorm}
O_g^B(z) \to Z_{gg}(z) O_{g}^R(z) , \ \ \  
O_q^B(z)\to       Z_{qq}(z) O_{q}^R(z) ,
\end{align}
where the superscript $B(R)$ denotes the bare(renormalized) operator. The counterterm behaves as~\cite{Wang:2019tgg}
\begin{equation}
\label{12}
   Z_{ii}(z)=e^{\delta m_i|z|}Z_{ii}(0) 
\end{equation}
without summing over the repeated indexes. When $z=0$, the right hand sides of Eq.(\ref{renorm}) could have more terms arising from the $gq$ mixing under renormalization, similar to the mixing of quark and gluon energy momentum tensor operators under renormalization.  
But when $z\ne 0$, these terms are negligible compared to the terms in Eq.(\ref{renorm}) in the continuum limit. The difference between the $z=0$ and $z\ne 0$ cases is that there is an additional Wilson line in the latter case, which behaves like a heavy quark propagator with an inverse power of momentum in the auxiliary field approach and makes the loop diagram more convergent than the former case.
Note that the $O^B$ and $O^R$ operators could have different structures, just as $O^B_q$ and $O^R_q$ in our case. 
 }


\subsection{The ratio scheme}
\label{r-scheme}

\blueX{The renormalization property in Eq.(\ref{renorm})} allows for non-perturbative renormalization (NPR). In the ratio scheme \Red{\cite{Radyushkin:2017cyf, Izubuchi:2018srq}}, the renormalized matrix element is 
\begin{equation}
\label{ratio}
  \frac{h_{i}^{r}(\zeta,z^2)}{\langle x \rangle_i} \equiv \frac{h_{i}^{B}(\zeta,z^2)}{h_{i}^{B}(0,z^2)}=\frac{Z_{ii}(z)h^R_i(\zeta,z^2)}{Z_{ii}(z)h^R_i(0,z^2)}=\frac{h^R_i(\zeta,z^2)}{h^R_i(0,z^2)} ,
\end{equation}
which is UV finite and where $\langle x \rangle_i$ is the second moment of the lightcone PDF for parton $i$. \blue{The function $h_{i}(\zeta,z^2)$ is defined in Eq.(\ref{h-functionX}) near the $P^z \to \infty$ limit. The extraction of $h_{i}(0,z^2)$ requires more care and is discussed in Appendix \ref{App0}.} 

\blueX{
In Eq.(\ref{ratio}), NPR only requires the ratio $h_{i}^{B}(\zeta,z^2)/h_{i}^{B}(0,z^2)$ to be finite in the continuum limit, it does not matter that $O^B_q$ and $O^R_q$ have different structures. 
What is important is that in the factorization theorem of Eq.(\ref{OpRenorm})
$h_{i}^{B}(\zeta,z^2)$ has the same infrared(IR) behavior as the PDF part (i.e. the $y$ integral). It does not matter what the ratio $h_{i}^{B}(\zeta,z^2)/h_{i}^{B}(0,z^2)$ is because no matter what the NPR subtraction is in the ultraviolet(UV), its effect will be canceled by the matching kernel $\mathcal{C}_{ij}^B$.   
} 

The quasi-PDF in this scheme can be obtained by the Fourier transformation of Eq.(\ref{Fourier}), which, after integrating over $x$, yields
\begin{equation}
\label{momconX}
\langle x \rangle_{\tilde{i}} = \langle x \rangle_i ,
\end{equation}
where $\langle x \rangle_{\tilde{i}} $ is the second moment of the quasi-PDF for parton $i$, with $i=u,d,s,g$ because the arguments that lead to Eq.(\ref{momconX}) not only apply to singet PDFs but also to non-singlet ones as well. However, in our notion, the support of PDFs is $x=[-1,1]$, which is related to the notion of global analysis (GA) with support $x=[0,1]$ as
\begin{align}
\langle x \rangle_{q}=& \langle x \rangle^{GA}_q+\langle x \rangle^{GA}_{\bar{q}}\nonumber \\
\langle x \rangle_{g}=&2 \langle x \rangle^{GA}_g
\end{align}
Therefore, momentum conservation 
\begin{equation}
\label{momcon}
\frac{1}{2}\langle x \rangle_{\tilde{g}}+\sum_{\tilde{q}_i}\langle x \rangle_{\tilde{q}_i}= \frac{1}{2}\langle x \rangle_{g}+\sum_{q_i}\langle x \rangle_{q_i} =1 
\end{equation}
for quasi-PDFs in the ratio scheme is built-in.


For a nucleon moving with momentum $P^z$ that is much larger than the nucleon mass $M$ and $\Lambda_{\text{QCD}}$, the quasi-PDF can be related to PDF through a factorization theorem \Red{\cite{Izubuchi:2018srq}}. 
In coordinate space, we have the factorization 
\begin{align}
\label{OpRenorm}
    h_i^B(\zeta,z^2)
    =\sum_j\int^1_{-1}d\alpha\,\mathcal{C}_{ij}^B(\alpha,z^2)\int^1_{-1}dy\,e^{-i\alpha y\zeta} y f_j^{\msb}(y)
    +\mathcal{O}(z^2 M^2,z^2\Lambda_{QCD}^2) ,
\end{align}
where we have replaced the renormalized correlator $h_i^R$ in Ref. \cite{Izubuchi:2018srq} by the bare one $h_i^B$ such that the UV regulator dependence 
of the quasi-PDF in $h_i^B$ and $\mathcal{C}_{ij}$ remains unsubtracted (while the lightcone PDF $f_j$ never depends on this regulator). The separation of UV and IR physics still holds despite this replacement. The matching coefficient $\mathcal{C}_{ij}$ compensates the difference in the UV physics between the quasi-PDF and PDF. It can be computed in perturbative QCD and is free of infrared singularity. 
Up to $\mathcal{O}(z^2)$ corrections,  
\begin{align}
\label{CT}
    h_i^B(0,z^2)
    =\sum_j\int^1_{-1}d\alpha\,\mathcal{C}_{ij}^B(\alpha,z^2) \langle x \rangle_j
    =\langle x \rangle_i\int^1_{-1}d\alpha\,\left[\mathcal{C}_{ii}^B(\alpha,z^2)+\sum_{j \ne i}\mathcal{C}_{ij}^B(\alpha,z^2) \frac{\langle x \rangle_j}{\langle x \rangle_i}\right] .
\end{align} 

We can plug Eqs.(\ref{OpRenorm}) and (\ref{CT}) into Eq.(\ref{ratio}), 
\begin{align}
    h_i^r(\zeta,z^2)
    &=\sum_j\int^1_{-1}d\alpha\,\mathcal{C}_{ij}(\alpha,z^2)\int^1_{-1}dy\,e^{-i\alpha y\zeta} y f_j^{\msb}(y)
    +\mathcal{O}(z^2 M^2,z^2\Lambda_{QCD}^2) ,
\end{align}
where
\begin{align}
\label{20}
    \mathcal{C}_{ij}(\alpha,z^2)&=\frac{\mathcal{C}_{ij}^B(\alpha,z^2)}{\int^1_{-1}d\alpha\,\left[\mathcal{C}_{ii}^B(\alpha,z^2)+\sum_{j \ne i}\mathcal{C}_{ij}^B(\alpha,z^2) \frac{\langle x \rangle_j}{\langle x \rangle_i}\right]}
\end{align}
which can be expanded in $\alpha_s$ using $\mathcal{C}_{ij}=\delta(\alpha-1)\delta_{ij}+\mathcal{C}^{(1)}_{ij}$ with $\mathcal{C}^{(1)}_{ij}=\mathcal{O}(\alpha_s)$.

The Fourier transform of $h_i^r(z^2,\zeta)$ using Eq.(\ref{Fourier}) yields the factorization formula in momentum space: 
\begin{equation}
\label{factorization}
x\tilde{f}^r_i(x,P^z)=\sum_j\int_{-1}^{1}\frac{dy}{|y|}C_{ij}(\frac{x}{y},\frac{\mu}{yP^z})yf_j(y,\mu)+\mathcal{O}\left(\frac{M^2}{(P^z)^2},\frac{\Lambda_{QCD}^2}{(P^z)^2}\right) ,
\end{equation}
where we have used
the relation between the coordinate space and momentum space matching kernels 
\begin{align}
    C_{ij}(\frac{x}{y},\frac{\mu}{yP^z})
    =\int_{-\infty}^{\infty}\frac{d\zeta}{2\pi}e^{i\zeta \frac{x}{y} }\int^1_{-1}d\alpha\,e^{-i\alpha \zeta}\mathcal{C}_{ij}(\alpha,\frac{\zeta^2}{(y P^z)^2}) .
    \label{MK_mo}
\end{align}

The factorization formula of Eq.(\ref{factorization}) can be straightforwardly generalized to non-singlet quarks as well such that the index $i(j)$ spans over $\{g,u,d,s\}\equiv \{g,q_{i(j)}\}$. Here we have focused on light (u,d, and s) quark PDFs only hence the number of quark flavors $n_f=3$. Also, we will take the $m_u=m_d=m_s=0$ limit in the matching kernel. In this limit, $C_{gq_i}$ and $C_{q_iq}$ are independent of the quark flavor $i$,  
\begin{equation}
     C_{gq_i}=C_{gq}, \ \ \ C_{q_ig}=C_{qg} .
\end{equation}
And no flavor changing in QCD implies 
\begin{equation}
     \ \Red{C_{q_iq_j}=C_{q_iq_i}\delta_{ij}}
\end{equation}
where the $i$ index is not summed over. \Red{So, we have the full matching formula:} 

\begin{equation}
\label{matchingX}
    \begin{split}
    x\tilde{g}^r(x,P^z)&=\int_{-1}^{1}\frac{dy}{|y|}\left[C_{gg}(\frac{x}{y},\frac{\mu}{yP^z})yg(y,\mu)+C_{gq}(\frac{x}{y},\frac{\mu}{yP^z})y\sum_{j=u,d,s}q_j(y,\mu) \right]+\mathcal{O}\left(\frac{M^2}{(P^z)^2},\frac{\Lambda_{QCD}^2}{(P^z)^2}\right) ,\\
    x\tilde{q}_{i}^r(x,P^z)&=\int_{-1}^{1}\frac{dy}{|y|}\left[\Blue{C_{q_iq_i}}(\frac{x}{y},\frac{\mu}{yP^z})yq_{i}(y,\mu)+C_{qg}(\frac{x}{y},\frac{\mu}{yP^z})yg(y,\mu) \right]+\mathcal{O}\left(\frac{M^2}{(P^z)^2},\frac{\Lambda_{QCD}^2}{(P^z)^2}\right) .
    \end{split}
\end{equation}
As we will show later, $C_{gg}$, $C_{gq}$, and $C_{qg}$ are even functions of $x/y$ \Blue{(due to crossing symmetry of the Feynman diagrams)} and $\mu/(yP^z)$, and $g(y,\mu)$ is odd in $y$ due to charge conjugation symmetry (the number of gluon equals the number of anti-gluon). Therefore, only the $y$ odd part of $q_i(y,\mu)$
mixes with $g(y,\mu)$ in the matching formula to give rise to $\tilde{q}_i(x,P^z)$ and $\tilde{g}(x,P^z)$ that are odd in $x$. The $y$ even part of $q_i(y,\mu)$ does not mix with $g(y,\mu)$ in the matching. It gives rise to the even in $x$ part of 
$\tilde{q}_i(x,P^z)$.

In Appendix \ref{AppA}, we 
show how the one-loop matching factors in Eq.(\ref{matchingX})
are computed.
Here we first present the matching between the lightcone PDFs in the $\overline{\text{MS}}$ scheme and the quasi-PDFs in the ratio scheme, then change the ratio scheme to the hybrid-ratio scheme \Blue{(defined later)} in Sec. \ref{H-r}:

\begin{equation}
    \begin{split}
        \Red{C_{q_iq_i}}(x,\frac{\mu}{p^z})&=\delta(x-1) \\
        &+\frac{\alpha_sC_F}{\Red{2}\pi}\left\{ \begin{array}{rcl} &\left[x\frac{1+x^2}{1-x}\mathrm{ln}(\frac{x}{-1+x})+x+\frac{3}{2}+\frac{17}{6}\frac{1}{x-1}\right]_{+(1)}^{[1,\infty]} & \mbox{for}\ 1<x \\ &\left[x\frac{1+x^2}{1-x}(-\mathrm{ln}(\frac{\mu^2}{4(1-x)x p_z^2}))-\frac{x^2(1+x)}{1-x}+\frac{17}{6}\frac{1}{1-x}+\frac{3}{2}\right]_{+(1)}^{[0,1]} & \mbox{for}\  0<x<1 \\ &\left[-x\frac{1+x^2}{1-x}\mathrm{ln}(\frac{-x}{1-x})-x-\frac{3}{2}+\frac{17}{6}\frac{1}{1-x}\right]_{+(1)}^{[-\infty,0]} & \mbox{for}\ x<0 \end{array}\right.\\
        &-\frac{\alpha_sT_F}{\Red{2}\pi}\left\{ \left[\frac{1}{3}-\frac{1}{3}\mathrm{ln}\left(\frac{\mu^2}{4p_z^2}\right)\right]\delta(1-x)+\frac{1}{3}\left[\left[\frac{1}{|1-x|} \right]_{+(1)}^{[0,2]}+\frac{1}{|x-1|}\theta(-x)+\frac{1}{|1-x|}\theta(x-2) \right] \right\}\frac{\langle x\rangle_g}{\Red{\langle x\rangle_i}} ,
    \end{split}
    \label{ratio_match_qq_1}
\end{equation}
where $C_F=(N^2-1)/(2N)$, $C_A=N$, with the number of colors $N=3$, and $T_F=1/2$. The plus function has the property
\begin{equation}
    \int_{-\infty}^{\infty}dx\ \left[ f(x) \right]_{+(c)}^{[a,b]}g(x)=\int_{a}^{b}dx\ f(x)\left[ g(x)-g(c) \right] .
\end{equation}
\begin{equation}
    \begin{split}
        C_{gg}(x,\frac{\mu}{p^z}) &= \delta(x-1) \\
        &+\frac{\alpha_s C_A}{4\pi}\left\{ \begin{array}{rcl} & \left[\frac{2(1-x+x^2)^2}{-1+x}\mathrm{ln}(\frac{x-1}{x})+2x^2-x+\frac{11}{3}+\frac{11}{6}\frac{1}{x-1}\right]_{+(1)}^{[1,\infty)} & \mbox{for}\ 1<x \\ & \left[\frac{2(1-x+x^2)^2}{-1+x}\left(\mathrm{ln}(\frac{\mu^2}{4(1-x)x p_z^2})\right)+\frac{29-30x+12x^2-12x^3}{6(1-x)}+\frac{2}{3}\right]_{+(1)}^{[0,1]} & \mbox{for}\  0<x<1 \\ &(-1)\left[\frac{2(1-x+x^2)^2}{-1+x}\mathrm{ln}(\frac{x-1}{x})+2x^2-x+\frac{11}{3}+\frac{11}{6}\frac{1}{x-1}\right]_{+(1)}^{(-\infty,0]} & \mbox{for}\ x<0 \end{array}\right.\\
        &-\frac{\alpha_sC_F}{\Blue{2}\pi}\left\{ \left[\frac{1}{3}-\frac{4}{3}\mathrm{ln}\left(\frac{\mu^2}{4p_z^2}\right)\right]\delta(1-x)+\frac{4}{3}\left[\left[\frac{1}{|1-x|} \right]_{+(1)}^{[0,2]}+\frac{1}{|x-1|}\theta(-x)+\frac{1}{|1-x|}\theta(x-2) \right] \right\}\frac{\sum_{i=u,d,s}\langle x\rangle_i}{\langle x\rangle_g}\\
        &+\{x\rightarrow -x\} .
    \end{split}
    \label{ratio_match_gg}
\end{equation}
\begin{equation}
\label{gq}
    \begin{split}
        C_{gq}(x,\frac{\mu}{p^z}) &= \frac{\alpha_sC_F}{\Blue{2}\pi}\left\{ \begin{array}{rcl} & \left[(2-2x+x^2)\mathrm{ln}(\frac{x}{-1+x})-x+\frac{3}{2}-\frac{4}{3}\frac{1}{x-1}\right]_{+(1)}^{[1,\infty)} & \mbox{for}\ 1<x \\ & \left[x+x^2-(2-2x+x^2)\mathrm{ln}(\frac{\mu^2}{4(1-x)x p_z^2})-\frac{4}{3}\frac{1}{1-x}\right]_{+(1)}^{[0,1]} & \mbox{for}\  0<x<1 \\ &(-1)\left[(2-2x+x^2)\mathrm{ln}(\frac{x}{-1+x})-x+\frac{3}{2}-\frac{4}{3}\frac{1}{x-1}\right]_{+(1)}^{(-\infty,0]} & \mbox{for}\ x<0 \end{array}\right.\\
        &+\frac{\alpha_sC_F}{\Blue{2}\pi}\left\{ \left[\frac{1}{3}-\frac{4}{3}\mathrm{ln}\left(\frac{\mu^2}{4p_z^2}\right)\right]\delta(1-x)+\frac{4}{3}\left[\left[\frac{1}{|1-x|} \right]_{+(1)}^{[0,2]}+\frac{1}{|x-1|}\theta(-x)+\frac{1}{|1-x|}\theta(x-2) \right] \right\}\\
        &+\{x\rightarrow -x\} .
    \end{split}
\end{equation}
\begin{equation}
\label{qg}
    \begin{split}
        C_{qg}(x,\frac{\mu}{p^z}) &= \frac{\alpha_s T_F}{4\pi}\left\{ \begin{array}{rcl} &\left[x(1-2x+2x^2)\mathrm{ln}(\frac{x}{-1+x})-2x^2+x-\frac{1}{3}\frac{1}{x-1}-\frac{2}{3}\right]_{+(1)}^{[1,\infty)} & \mbox{for}\ 1<x \\ & \left[\frac{5}{2}x-x^3-x(1-2x+2x^2)\mathrm{ln}(\frac{\mu^2}{4(1-x)x p_z^2})-\frac{1}{3}\frac{1}{1-x}-\frac{2}{3}\right]_{+(1)}^{[0,1]} & \mbox{for}\  0<x<1 \\ &(-1)\left[x(1-2x+2x^2)\mathrm{ln}(\frac{x}{-1+x})-2x^2+x-\frac{1}{3}\frac{1}{1-x}-\frac{2}{3}\right]_{+(1)}^{(-\infty,0]} & \mbox{for}\ x<0 \end{array}\right.\\
        &+\frac{\alpha_sT_F}{4\pi}\left\{ \left[\frac{1}{3}-\frac{1}{3}\mathrm{ln}\left(\frac{\mu^2}{4p_z^2}\right)\right]\delta(1-x)+\frac{1}{3}\left[\left[\frac{1}{|1-x|} \right]_{+(1)}^{[0,2]}+\frac{1}{|x-1|}\theta(-x)+\frac{1}{|1-x|}\theta(x-2) \right] \right\}\\
        &+\{x\rightarrow -x\} .
    \end{split}
\end{equation}
The $x \to -x$ terms arising from the crossed symmetry in $C_{gg}$, $C_{gq}$, and $C_{qg}$. Also, the infrared (IR) singularities, described by $1/\epsilon_{IR}$ poles, are canceled between quasi-PDFs and PDFs. The ultraviolet (UV) singularities, described by $1/\epsilon_{UV}$ poles, are canceled between quasi-PDFs and their counterterms defined in Eq.(\ref{ratio}).

\subsection{The hybrid-ratio scheme}
\label{H-r}

\Blue{In the ratio scheme, the counterterm $h_{i}^{B}(0,z^2)$ will have undesirable non-perturbative infrared(IR) contribution when $z>1/\Lambda_{\text{QCD}}$. The hybrid-renormalization scheme \Red{\cite{Ji:2020brr,Chou:2022drv}} fixes this problem by employing a different renormalization scheme at large $z$ without the IR physics subtraction. The simplest option is to use the counterterms of the form of Eq. (\ref{12}) at large $z$ such that
\begin{equation}
h_{i}^{hyb-r}(\zeta,z^2)= \frac{h_{i}^{B}(\zeta,z^2)}{Z^{hyb-r}(z)} ,
\end{equation}
with
\begin{equation}
    Z^{hyb-r}(z)=\frac{\theta(z_s-|z|)}{\langle x \rangle_i}\Red{h_{i}^{B}(0,z^2)}+e^{\delta m_i|z|}Z_{ii}(0)\theta(|z|-z_s) .
\end{equation}
The ratio scheme results of Eqs.(\ref{ratio})-(\ref{momcon}) in short distance still hold in the hybrid-ratio scheme. } Now the matching kernel will depend on the scale $z_s$ 
of the scheme change: 
\begin{equation}
\label{32}
    C_{ij}^{hyb-r}(x,\frac{\mu}{p^z},z_s p^z)=C_{ij}(x,\frac{\mu}{p^z})+\delta C_{ij}(x,z_s p^z) ,
\end{equation}
where $C_{ij}$ is the matching factor in the ratio scheme shown in Eqs.(\ref{ratio_match_qq_1})-(\ref{qg}). And
\begin{equation}
\label{33X}
    \delta C_{gg}(x,z_s p^z)=\left(-\frac{11\alpha_s C_A}{24\pi}+\frac{2\alpha_sC_F}{3\pi}\frac{\Red{\sum_{i}\langle x \rangle_{q_i}}}{\langle x \rangle_g}\right)\left[ \frac{1}{|x-1|}-\frac{2 \mathrm{Si}((1-x)z_s p^z)}{\pi(1-x)} \right]^{[-\infty,\infty]}_{+(1)}+\{x\rightarrow -x\} ,
\end{equation}
where
\begin{equation}
    \mathrm{Si}(x)=\int_{0}^{x}\frac{\sin(t)}{t}dt ,
\end{equation}
\begin{equation}
\label{35X}
    \delta \Red{C_{q_iq_i}}(x,z_s p^z)=\left(-\frac{17\alpha_s C_F}{24\pi}+\frac{\alpha_sT_F}{12\pi}\frac{\langle x \rangle_g}{\Red{\langle x \rangle_{q_i}}}\right)\left[ \frac{1}{|x-1|}-\frac{2 \mathrm{Si}((1-x)z_s p^z)}{\pi(1-x)} \right]^{[-\infty,\infty]}_{+(1)}
\end{equation}
and
\begin{equation}
    \delta C_{qg}(x,z_s p^z)=\delta C_{gq}(x,z_s p^z)=0.
\end{equation}
Up to $\mathcal{O}\left( M^2/(P^z)^2,\Lambda_{QCD}^2/(P^z)^2\right)$ corrections, 
\begin{equation}
\label{matchingXXX}
    \begin{split}
    x\tilde{g}^{hyb-r}(x,P^z,z_s)&=\int_{-1}^{1}\frac{dy}{|y|}\left[C^{hyb-r}_{gg}(\frac{x}{y},\frac{\mu}{yP^z}, y z_s P^z)yg(y,\mu)+C^{hyb-r}_{gq}(\frac{x}{y},\frac{\mu}{yP^z}, y z_s P^z)y\sum_{j=u,d,s}q_j(y,\mu) \right] ,\\
    x\tilde{q}_{i}^{hyb-r}(x,P^z,z_s)&=\int_{-1}^{1}\frac{dy}{|y|}\left[C^{hyb-r}_{\Red{q_iq_i}}(\frac{x}{y},\frac{\mu}{yP^z}, y z_s P^z)yq_{i}(y,\mu)+C^{hyb-r}_{qg}(\frac{x}{y},\frac{\mu}{yP^z}, y z_s P^z)yg(y,\mu) \right] ,
    \end{split}
\end{equation}
where $i=u,d,s$. One can show that both $x\tilde{g}^{hyb-r}(x)$ and $x\tilde{q}_{i}^{hyb-r}(x)$ behave as $1/x^2$ when $x\to \pm \infty$ through the one loop matching. Also, the ``momentum conservation" relation
\begin{align}
\label{MC}
\int dx\ x\tilde{g}^{hyb-r}(x,P^z,z_s)=\int dx\ x g(x,\mu) , \nonumber\\  \int dx\ x\tilde{q}_i^{hyb-r}(x,P^z,z_s)=\int dx\ x q_i(x,\mu) , \
\end{align}
is proved at one loop in the Appendix~\ref{App3}. 
\Blue{This behavior is similar to what was seen in Ref.~\cite{Chou:2022drv} for non-singlet quark PDF in the hybrid-ratio scheme. The UV divergence together with the short distance 
$\ln z$ dependence cancels after forming the ratio. Therefore, local vector current conservation is reached smoothly by taking the ratio of correlators of finite $z$ to the $z\to 0$ limit, while the quark quasi-PDF $\tilde{q}(x)$ 
behaves as $1/x^2$ as $|x| \to \infty$. In contrast, the same quasi-PDF correlator in the $\msb$ scheme has short distance $\ln z$ dependence hence the $z\to 0$ limit to \Red{recover} the conserved current result at $z=0$ is not smooth. This implies sensitivity to short distance physics
with the corresponding $\tilde{q}(x)$ 
behaves as $1/x$ as $|x| \to \infty$ with the current conservation formally recovered by delta functions at infinity $|x|$. 
}

\subsection{\Blue{Comment on fermion number conservation}}
    While Eq.(\ref{matchingXXX}) satisfies the ``momentum conservation" defined in Eq.(\ref{MC}), it can be shown that the fermion number conservation 
\begin{align}
\label{PNC}
\int dx\ \tilde{q_i}^{hyb-r}(x,P^z,z_s)= \int dx\ q_i(x)
\end{align}
is not satisfied. This is because 
the quark operator in Eq.(\ref{Oq}) does not
yield a vector current operator in the $z\to 0$ limit. In fact, the operator product expansion of the quark operator 
does not include the vector current operator at all. In general, this should not be a problem. Because all we need is fermion number conservation on the extracted PDFs, not the quasi-PDFs.

\begin{figure}[tbp]
\centering
\includegraphics[width=1.\textwidth, height=0.3\textwidth]{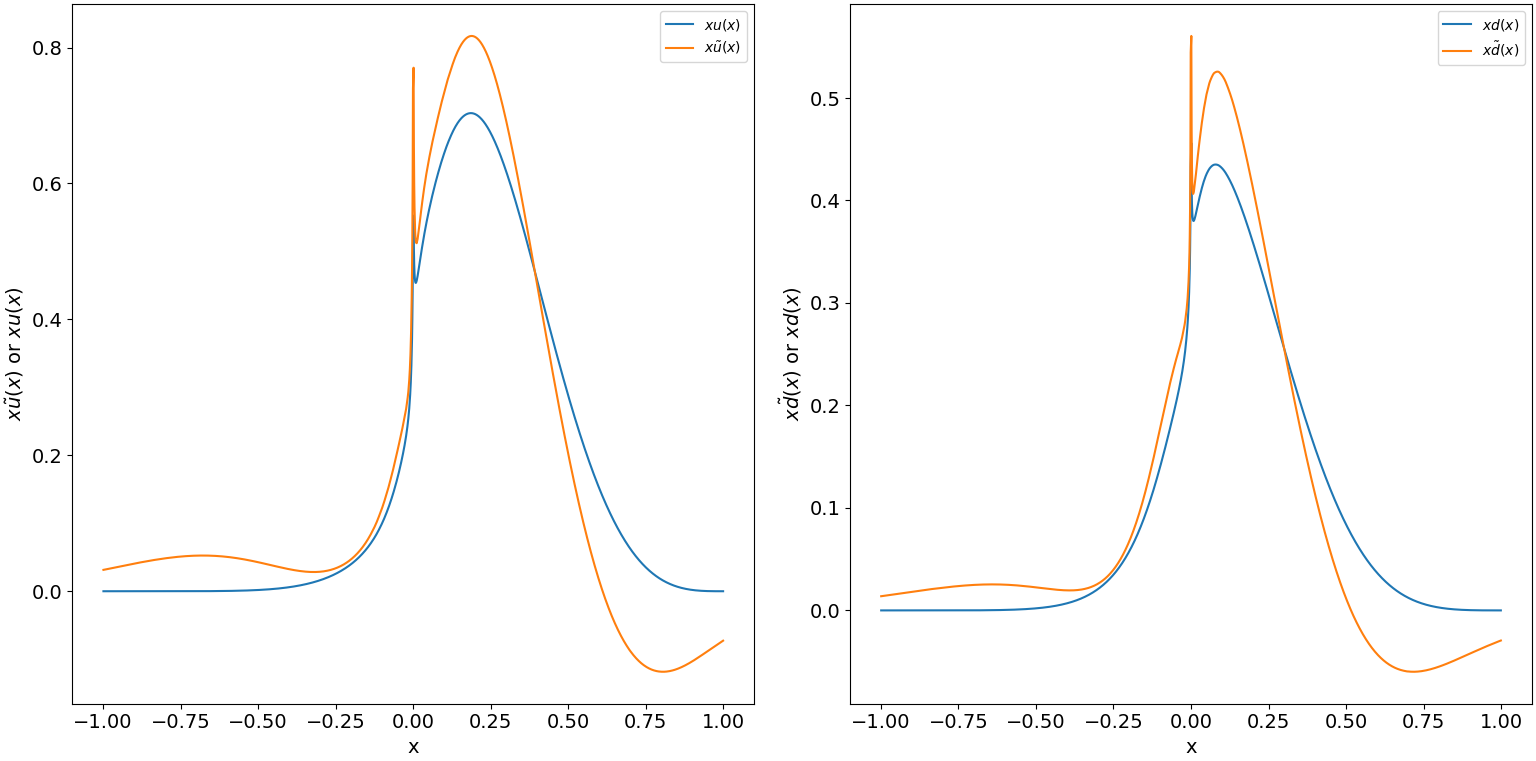}
\ \
\ \
\includegraphics[width=1.\textwidth, height=0.3\textwidth]
{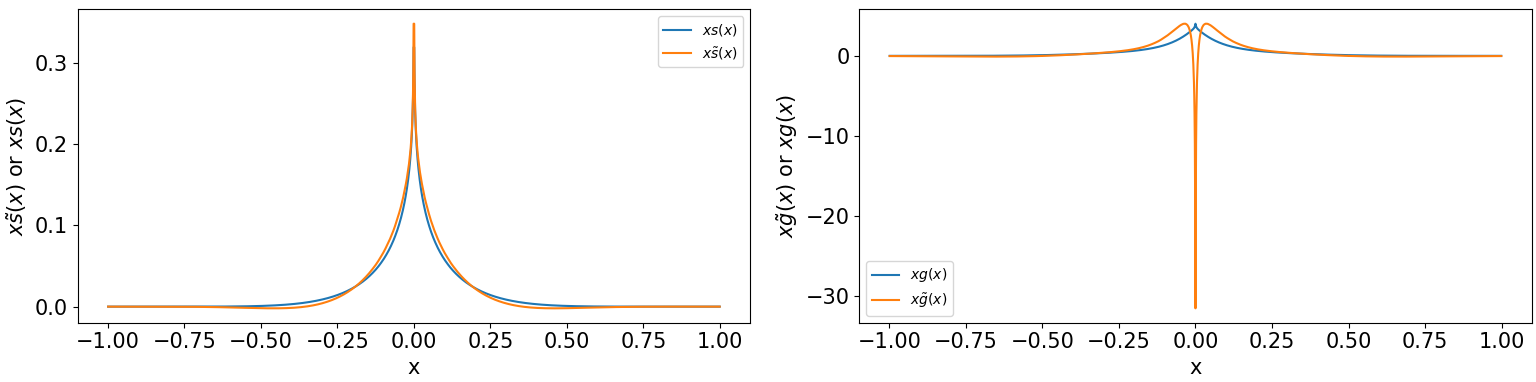}

\caption{ Quasi-PDFs for $xu$, $xd$, $xs$ and $xg$ obtained using the matching formula of Eq.(\ref{matchingXXX}) but with the modification of Eq.(\ref{new_match}). The inputs are the CT14 proton PDFs~\cite{Dulat_2016} with $\mu=2.0$ GeV,   
$\alpha_s=0.283$, $P_z=1.0$ GeV, and $z_s=0.3$. }
\label{fig:full_alter}
\end{figure}

\begin{figure}[tbp]
\centering
\includegraphics[width=1.\textwidth, height=0.3\textwidth]{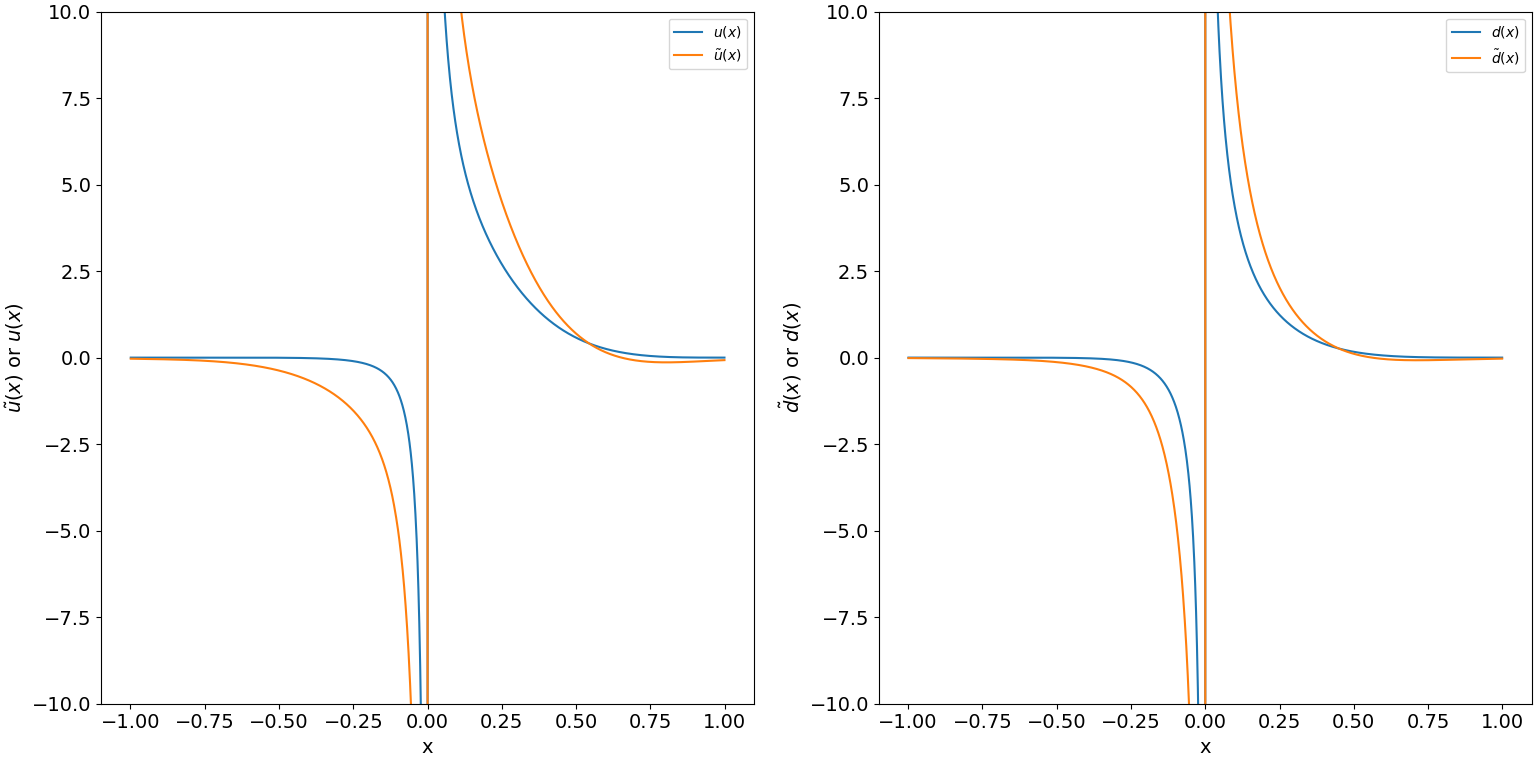}
\caption{Quasi-PDFs obtained using the matching formula of Eq.(\ref{matchingXXX}) but with the modification of Eq.(\ref{new_match}). $\int dx\ s(x)=\int dx\ \tilde{s}(x)=\int dx\ g(x)=\int dx\ \tilde{g}(x)=0$ because they are odd functions in $x$. The inputs are the same as Fig.(\Ref{fig:full_alter}).}
\label{fig:all2x}
\end{figure}

However, one can also impose both fermion number and momentum conservation for the quasi-PDFs by using a different set of quark operators: 
\begin{align}
O_{q,O}(z)=
\frac{-i}{2}\frac{d}{dz}\left[\bar{\psi} (z)\gamma^z U(z,0)\psi(0)-\bar{\psi} (-z)\gamma^z U(-z,0)\psi(0)\right], \\
O_{q,E}(z)=
\frac{1}{2}\left[\bar{\psi} (z)\gamma^z U(z,0)\psi(0)+\bar{\psi} (-z)\gamma^z U(-z,0)\psi(0)\right] ,
\end{align}
\Blue{The denominator in the ratio scheme for using $O_{q,O}$ and $O_q$ will be different. 
But the difference is of higher order ($\mathcal{O}(\alpha_s^2)$).}
$O_{q,O(E)}$ generates a quasi-PDF that is odd(even) in $x$. 
The gluon operator $O_{g}$ generates a gluon quasi-PDF which is odd in $x$ due to charge conjugation symmetry. $O_{g}$ only mixes with $O_{q,O}$ but not $O_{q,E}$. (Note that $O_{q,O}$ and $O_{q,E}$ are of different dimensions, therefore the corresponding dimensionless correlator for $O_{q,E}$ is $\langle P|O_{q,E}(z)|P\rangle/2 P^z$ instead of Eq.(\ref{h-functionX})). Then the $C_{q_iq_i}$ term in the matching formula of \Red{Eq.(\ref{matchingXXX})} becomes 
\begin{equation}
C_{q_iq_i}^{hyb-r}(\frac{x}{y})yq_i(y) \to C_{q_iq_i}^{hyb-r}(\frac{x}{y})yq_{i,O}(y)+\bar{C}_{q_iq_i}^{hyb-r}(\frac{x}{y})x q_{i,E}(y) ,
\label{new_match}
\end{equation}
where $q_{i,O(E)}(y)=[q_{i}(y)\mp q_{i}(-y)]/2$ and
\begin{align}
&\bar{C}_{q_iq_i}^{hyb-r}\left(\xi,\frac{\mu}{y P^z},y z_sP^z\right)\\
=&\delta(1-\xi)+\frac{\alpha_sC_F}{2\pi}
    \begin{cases}
    \p{\frac{1+\xi^2}{1-\xi}\ln\frac{\xi}{\xi-1}+1}{1}{\infty}{1}&,\xi>1\\
    \p{\frac{1+\xi^2}{1-\xi}\left(-\ln\frac{\tilde{\mu}^2}{y^2P_z^2}+\ln4\xi(1-\xi)-1\right)+1+2(1-\xi)}{0}{1}{1}&,0<\xi<1\\
    \p{-\frac{1+\xi^2}{1-\xi}\ln\frac{-\xi}{1-\xi}-1}{-\infty}{0}{1}&,\xi<0
    \end{cases}\nonumber\\
    +&\frac{3\alpha_sC_F}{2\pi^2}\p{\frac{\text{Si}((1-\xi)
    \vert y \vert
    z_sP^z)}{(1-\xi)}}{-\infty}{\infty}{1}
\end{align}
\Red{is the matching kernel for non-singlet quark PDF in the hybrid-ratio scheme with $\xi\equiv x/y$.}

Using Eq.(\ref{new_match}), the quasi-PDFs and PDFs for $xu$, $xd$,$xs$ and $xg$  inside a proton are plotted in Fig.\ref{fig:full_alter}. The areas under the PDF and quasi-PDF curves in each plot are the same indicating that  Eqs.(\ref{MC})  are satisfied. 

The $s(x)$ and $g(x)$ are odd functions of $x$ with $\int dx\tilde{s}(x)=\int dx s(x)=\int dx\tilde{g}(x)=\int dx g(x)=0$ (meaning the numbers of partons and anti-partons are the same for $s$ and $g$ inside a proton.).

To see the numerical effect of \Blue{Eq.(\ref{matchingXXX}) but with the modification of Eq.(\ref{new_match})}, we use the proton  PDFs from CT14 global analysis~\cite{Dulat_2016} with $\mu=2.0$ GeV in the $\msb$ scheme,   
$\alpha_s=0.283$, $P_z=1.0$ GeV, and $z_s=0.3$. The resulting quasi-PDFs shown in Fig.(\ref{fig:full_alter})
are having areas under the curves the same as the area under the curves of the PDFs. This confirms that Eq.(\ref{MC}) is satisfied.

In Fig.(\Ref{fig:all2x}), 
$u(x)$ and $d(x)$ and their 
quasi-PDFs obtained using the matching formula of \Blue{Eq.(\ref{matchingXXX}) but with the modification of Eq.(\ref{new_match})} are ploted. 
The resulting quasi-PDFs shown in Fig.(\ref{fig:all2x})
are having areas under the curves the same as the area under the curves of the PDFs. This confirms that Eq.(\ref{PNC}) is satisfied. $\int dx s(x)=\int dx \tilde{s}(x)=\int dx g(x)=\int dx \tilde{g}(x)=0$ becasue they are odd functions in $x$ (meaning the numbers of partons and anti-partons are the same for $s$ and $g$ inside a proton).
The inputs are the same as Fig.(\Ref{fig:full_alter}).  

\subsection{\Blue{Singlet and Non-Singlet PDF Decomposition}}
\label{singlet}

In the previous section, the $\{u,d,s,g\}$ basis is used. In this basis, the quark-quark mixing kernels are not flavor blind, such that one cannot further separate the quark flavors into singlet and non-singlet PDFs, with the former mixing with the gluon PDF but the latter do not. Therefore, One has to work with the four sets of quasi-PDF's simultaneously in order to extract the corresponding PDF's. 

This problem can be fixed if we choose the flavor basis $q_i=\{q_3,q_8,q_S\}\equiv\{u-d,u+d-2s,u+d+s\}$ instead of the $\{u,d,s\}$ basis, then only $q_S$ mixes with the gluons. Hence, $\tilde{q}_3$ and $\tilde{q}_8$ are nonsinglet quasi-PDFs that will not mix with any other quasi-PDFs. $\tilde{q}_S$ and $\tilde{g}$ are singlet quasi-PDFs that will mix between them.

\begin{figure}[tbp]
\centering
\includegraphics[width=1.\textwidth, height=0.65\textwidth]{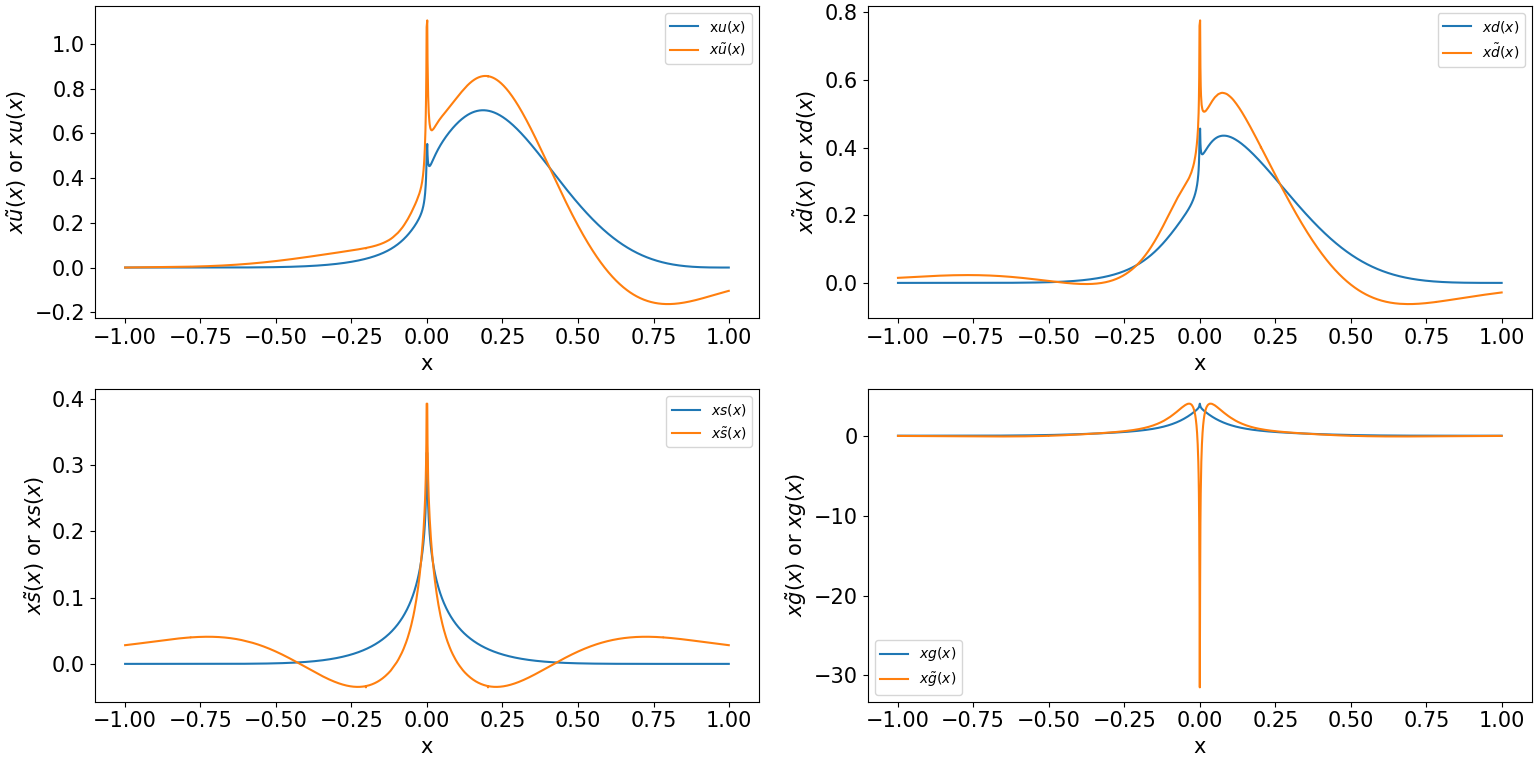}
\ \
\includegraphics[width=1.\textwidth, height=0.3\textwidth]{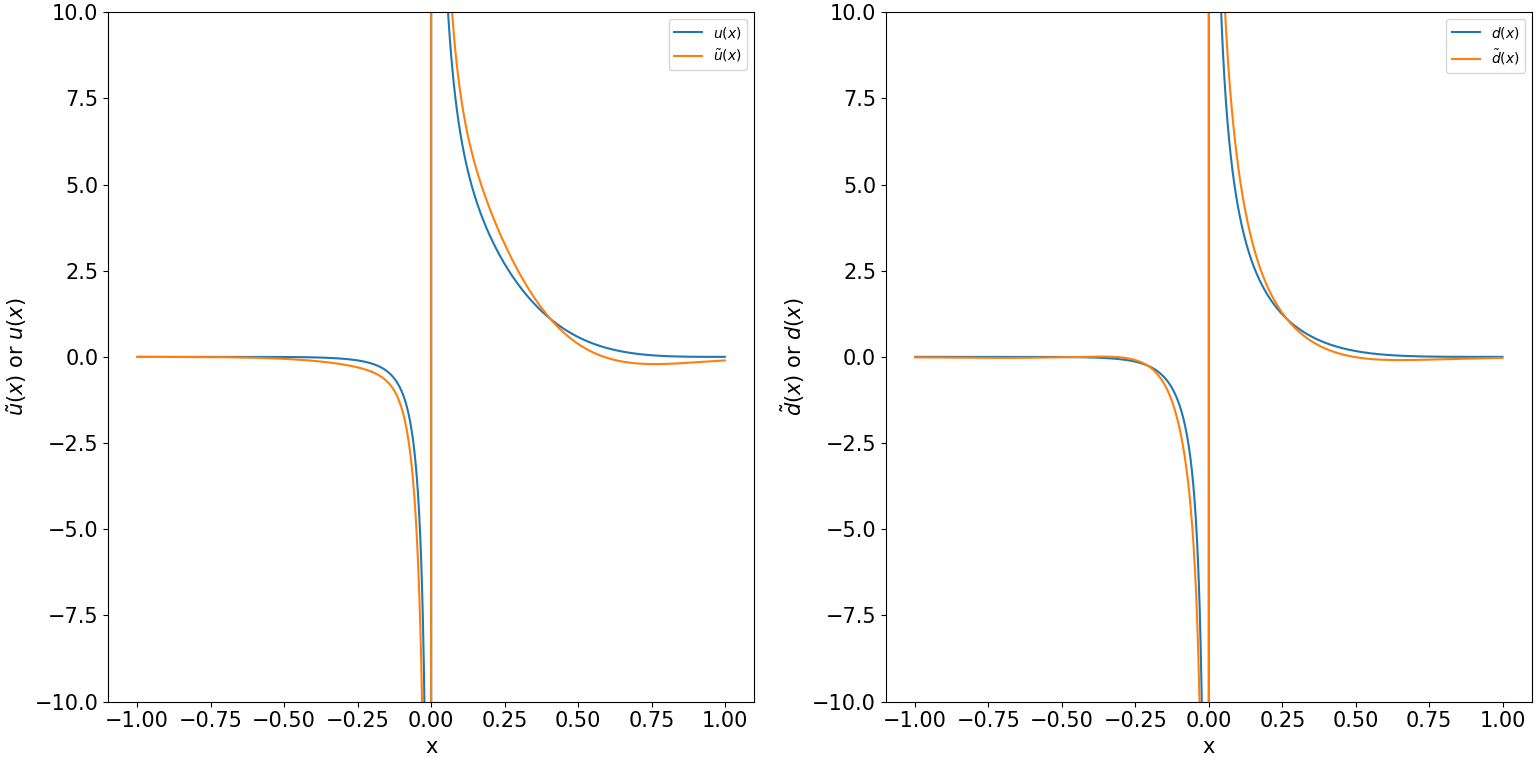}
\caption{Some inputs as Figs. \ref{fig:full_alter} and \ref{fig:all2x} but using the $\{q_3,q_8,q_S\}\equiv\{u-d,u+d-2s,u+d+s\}$ basis as described in Sec.\ref{singlet}. Only the singlet PDFs $q_S$ and $g$ mix with each other, the non-singlet PDFs $q_3$ and $q_8$ do not mix with any other PDF. We still plot the final result in the $\{u,d,s\}$ flavors.
}
\label{fig:3}
\end{figure}

In the following, we just list the equations that are changed under this new basis; otherwise, they remain the same. 
First, Eq.(\ref{matchingX}) becomes
\begin{equation}
\label{matchingXX}
    \begin{split}
    x\tilde{g}^r(x,P^z)&=\int_{-1}^{1}\frac{dy}{|y|}\left[C_{gg}(\frac{x}{y},\frac{\mu}{yP^z})yg(y,\mu)+C_{gq}(\frac{x}{y},\frac{\mu}{yP^z})yq_S(y,\mu) \right]+\mathcal{O}\left(\frac{M^2}{(P^z)^2},\frac{\Lambda_{QCD}^2}{(P^z)^2}\right) ,\\
    x\tilde{q}_{i}^r(x,P^z)&=\int_{-1}^{1}\frac{dy}{|y|}\left[\Blue{C_{q_iq_i}}(\frac{x}{y},\frac{\mu}{yP^z})yq_{i}(y,\mu)+\delta_{qq_S}C_{qg}(\frac{x}{y},\frac{\mu}{yP^z})yg(y,\mu) \right]+\mathcal{O}\left(\frac{M^2}{(P^z)^2},\frac{\Lambda_{QCD}^2}{(P^z)^2}\right) .
    \end{split}
\end{equation}
The matching kernel $C_{q_iq_i}$ of Eq. (\ref{ratio_match_qq_1}) is changed to
\begin{equation}
    \begin{split}
        \Red{C_{q_iq_i}}(x,\frac{\mu}{p^z})&=\delta(x-1) \\
        &+\frac{\alpha_sC_F}{\Red{2}\pi}\left\{ \begin{array}{rcl} &\left[x\frac{1+x^2}{1-x}\mathrm{ln}(\frac{x}{-1+x})+x+\frac{3}{2}+\frac{17}{6}\frac{1}{x-1}\right]_{+(1)}^{[1,\infty]} & \mbox{for}\ 1<x \\ &\left[x\frac{1+x^2}{1-x}(-\mathrm{ln}(\frac{\mu^2}{4(1-x)x p_z^2}))-\frac{x^2(1+x)}{1-x}+\frac{17}{6}\frac{1}{1-x}+\frac{3}{2}\right]_{+(1)}^{[0,1]} & \mbox{for}\  0<x<1 \\ &\left[-x\frac{1+x^2}{1-x}\mathrm{ln}(\frac{-x}{1-x})-x-\frac{3}{2}+\frac{17}{6}\frac{1}{1-x}\right]_{+(1)}^{[-\infty,0]} & \mbox{for}\ x<0 \end{array}\right.\\
        &-\delta_{q_iq_S}\frac{\alpha_sT_F}{\Red{2}\pi}\left\{ \left[\frac{1}{3}-\frac{1}{3}\mathrm{ln}\left(\frac{\mu^2}{4p_z^2}\right)\right]\delta(1-x)+\frac{1}{3}\left[\left[\frac{1}{|1-x|} \right]_{+(1)}^{[0,2]}+\frac{1}{|x-1|}\theta(-x)+\frac{1}{|1-x|}\theta(x-2) \right] \right\}\frac{\langle x\rangle_g}{\Red{\langle x\rangle_{q_i}}} .
    \end{split}
    \label{ratio_match_qq_1X}
\end{equation}
Eqs.(\ref{33X}) and (\ref{35X}) becomes
\begin{equation}
    \delta C_{gg}(x,z_s p^z)=\left(-\frac{11\alpha_s C_A}{24\pi}+\delta_{qq_S}\frac{2\alpha_sC_F}{3\pi}\frac{\Red{\langle x \rangle_{q_S}}}{\langle x \rangle_g}\right)\left[ \frac{1}{|x-1|}-\frac{2 \mathrm{Si}((1-x)z_s p^z)}{\pi(1-x)} \right]^{[-\infty,\infty]}_{+(1)}+\{x\rightarrow -x\} ,
\end{equation}
and
\begin{equation}
    \delta \Red{C_{q_iq_i}}(x,z_s p^z)=\left(-\frac{17\alpha_s C_F}{24\pi}+\delta_{q_iq_S}\frac{\alpha_sT_F}{12\pi}\frac{\langle x \rangle_g}{\Red{\langle x \rangle_{q_i}}}\right)\left[ \frac{1}{|x-1|}-\frac{2 \mathrm{Si}((1-x)z_s p^z)}{\pi(1-x)} \right]^{[-\infty,\infty]}_{+(1)} .
\end{equation}

\BlueX{The corresponding quasi-PDFs using the same PDF inputs as Figs. \ref{fig:full_alter} and \ref{fig:all2x} are shown in Fig.(\Ref{fig:3}) in the  $\{u,d,s\}$ flavor basis.
}

\end{widetext}

\section{Conclusions}

The one loop matching kernels between parton distribution functions (PDFs) for partons $i=u,d,s,g$ and their corresponding quasi-PDFs 
are computed at one loop in the hybrid-ratio scheme. We found that in addition to the conservation of the quasi-quark number for each flavor (as shown in Fig. \ref{fig:all2x}), 
the second moment $\langle x \rangle_{\tilde{i}}=\langle x \rangle_i$ of quasi-PDF and PDF of parton $i$ is the same in our approach (See Fig.\ref{fig:full_alter}). These are nice features of this approach. 


\section*{Acknowledgement}

JWC thanks  useful discussions with
Yushan  Su and Xingdong Ji during the LaMET2024 workshop that led to Eq.(12). This work is partly supported by the National Science and Technology Council, Taiwan, under Grants   112- 2112-M-002-027 and 113-2112-M-002-012. 
JWC thanks the InQubator for Quantum Simulation at the University of Washington and the Yukawa Institute for Theoretical Physics at Kyoto University for their hospitality.

\appendix

\begin{widetext}

\section{Extraction of $h_i(0,z^2)$ for Eq.(\ref{ratio})
}
\label{App0}

$h(\zeta, z^2)$ in Eq.(
\ref{h-functionX}) is defined near $P^z \to \infty$. However, $h(\zeta=z P^z=0, z^2)$ is needed in Eq.(\ref{ratio}). To extracty $h(0, z^2)$ we can start from the matrix element
\begin{align}
\label{Oden}
M^{\mu \nu}=
\langle P|F^{\{\mu\alpha}(\tilde{z})\tilde{U}(\tilde{z},0) F^{\ \nu\}}_{\alpha}(0)|P\rangle
, 
\end{align}
where the $\mu$ and $\nu$ indices enclosed by $\{ \dots \}$ are made symmetric traceless. The momentum of the external hadron is $P^{\mu}=(P^0,0,0,P^z)$ and the spacetime vector $\tilde{z}^{\mu}=(0,0,0,z)$. Since $P$ and $\tilde{z}$ are the only vectors in the problem, the matrix element can be decomposed to the following structure functions
\begin{align}
M^{\mu \nu}=
2 \left[h(\zeta, z^2) P^{\{\mu}P^{\nu\}}+k(\zeta, z^2) \tilde{z}^{\{\mu}\tilde{z}^{\nu\}}+l(\zeta, z^2) P^{\{\mu}\tilde{z}^{\nu\}}\right],
\end{align}
with the $\mu$ and $\nu$ indices symmetric and traceless. When $P^z \to \infty$, Eq.(\ref{h-functionX}) is reproduced. 

In the hadron rest frame, $P^{\mu}=(M,0,0,0)$, one obtains
\begin{align}
h(0, z^2) = \frac{M^{zz}+3 M^{xx}}{2 M^2} ,
\end{align}
which can be used in Eq.(\ref{ratio}).

Analogously, one can work with the matrix element of the quark operator
\begin{align}
M'^{\mu \nu}=
\langle P|-i\frac{d}{d\tilde{z}_{\{\mu}}\bar{\psi}_{q} (\tilde{z})\gamma^{\nu\}} U(\tilde{z},0)\psi_{q}(0)|P\rangle .
\end{align}
It is still the combination $(M'^{zz}+3 M'^{xx})/2 M^2$  that gives the desired structure function.

\section{One loop matching kernel computation}
\label{AppA}

In this Appendix, we show the result needed to obtain $\mathcal{C}_{ij}^B$ of Eq.(\ref{OpRenorm}). Once $\mathcal{C}_{ij}^B$ is obtained, the matching coefficient $C_{ij}$ of the ratio scheme can be determined using Eqs.(\ref{20}) and ( \ref{MK_mo}) and the matching coefficient $C_{ij}^{hybrid-r}$ of the hybrid-ration scheme can be obtained using Eq.(\ref{32}). 

The procedure to compute $\mathcal{C}_{ij}^B$ is described in Ref.~\cite{Radyushkin:2017cyf}. It starts with the matching formula between the psuedo-PDF defined as~\cite{Radyushkin:2017cyf}
\begin{equation}
\label{P1}
   \alpha \mathcal{P}^B_i(\alpha,z^2)\equiv \int_{-\infty}^{\infty}\frac{d\zeta}{2\pi} e^{i\alpha\zeta}h^B_{i}(\zeta,z^2)
\end{equation}
and the PDFs:
\begin{equation}
\label{P2}
    \alpha \mathcal{P}^B_i(\alpha,z^2)=\sum_j\int_{-1}^{1}\frac{dy}{|y|}\mathcal{C}^B_{ij}(\frac{\alpha}{y},z^2)yf_j^{\msb}(y,\mu).
\end{equation}
One can see Eq.(\ref{OpRenorm}) follows from Eqs.(\ref{P1}) and (\ref{P2}). Eq.(\ref{P2}) implies that
\begin{equation}
    \mathcal{C}_{ij}^{(1)^B}(\alpha,\mu^2z^2)=\mathcal{\bar{P}}^{(1)^B}_{i,j}(\alpha,\mu^2z^2)-\alpha f^{(1)^{\msb}}_{i,j}(\alpha),
\end{equation}
where $\mathcal{\bar{P}}^{(1)^B}_{i,j}$ is the one loop part of the pseudo-PDF of parton $i$ for a parton $j$ state. Similarly, $f^{(1)^{\msb}}_{i,j}$ is the one loop part of the lightcone PDF of parton $i$ for a parton $j$ state. The computation of $f^{(1)^{\msb}}_{i,j}$ is the same to the one that gives the DGLAP evolution kernel. Before renormalization, we have
\begin{align}
x f^{(1)}_{g,g}(x)=&\left\{\frac{\alpha_s C_A}{\Blue{4}\pi}\left[ \frac{2(1-x+x^2)^2}{1-x} \right]_{+(1)}^{[0,1]}-\frac{\alpha_s}{\Blue{4}\pi}\left[\frac{\Red{2}}{3}T_F n_f \right]\delta(1-x)\right\}\left( \frac{1}{\epsilon_{UV}}-\frac{1}{\epsilon_{IR}} \right)+ \{x\rightarrow -x\} ,\nonumber\\
x f^{(1)}_{q,g}(x)=&\frac{\alpha_sT_f}{4\pi}x(1-2x+2x^2)\left( \frac{1}{\epsilon_{UV}}-\frac{1}{\epsilon_{IR}} \right)\Red{\theta(x)\theta(1-x)}+\{x\rightarrow -x \} ,\nonumber\\
x f^{(1)}_{g,q}(x)=&
\frac{\alpha_sC_F}{\Red{2}\pi}(2-2x+x^2)\left( \frac{1}{\epsilon_{UV}}-\frac{1}{\epsilon_{IR}} \right)\Red{\theta(x)\theta(1-x)}+\{x \rightarrow -x \} ,\nonumber\\
x f^{(1)}_{q,q}(x)=&\frac{\alpha_sC_F}{2\pi}x\left[ \frac{1+x^2}{1-x} \right]^{[0,1]}_{+(1)}\left( \frac{1}{\epsilon_{UV}}-\frac{1}{\epsilon_{IR}} \right) ,
\end{align}
where the support $x=[-1,1]$ since we include the anti-gluon and anti-quark.

The computation of $\mathcal{\bar{P}}^{(1)^B}_{i,j}$ requires $h^B_{i,j}(z^2,\zeta)$ similar to the equal time correlator defined in
Eq.(\ref{h-functionX}) but with an eternal parton $j$ state:
\begin{equation}
\label{h-function}
h^B_{i,j}(p^z z,z^2)=\frac{1}{2 (p^z)^2}\langle p|O_i(z)|p\rangle_j ,
\end{equation}
with $i=g,q$. Then we can write
\begin{equation}
\label{h-function-II}
h^B_{i,j}(\zeta,z^2)=e^{-i \zeta}\delta_{ij}+h^{B,(1)}_{i,j}(\zeta,z^2)+\mathcal{O}(\alpha_s^2) .
\end{equation}
The result of $h^{B,(1)}_{i,j}(\zeta,z^2)$ is shown below. And since the ratio of Eq.(\ref{ratio}) is independent of the UV regulator. We will work with the dimensional regularization with regulator $\epsilon=(4-d)/2$ and separate  $\epsilon_{UV}>0$ and $\epsilon_{IR}<0$. We will work in the Feynman gauge.

\subsection{Gluon in gluon}
 
\begin{figure}[htp]
\begin{center}
\includegraphics[scale=0.5]{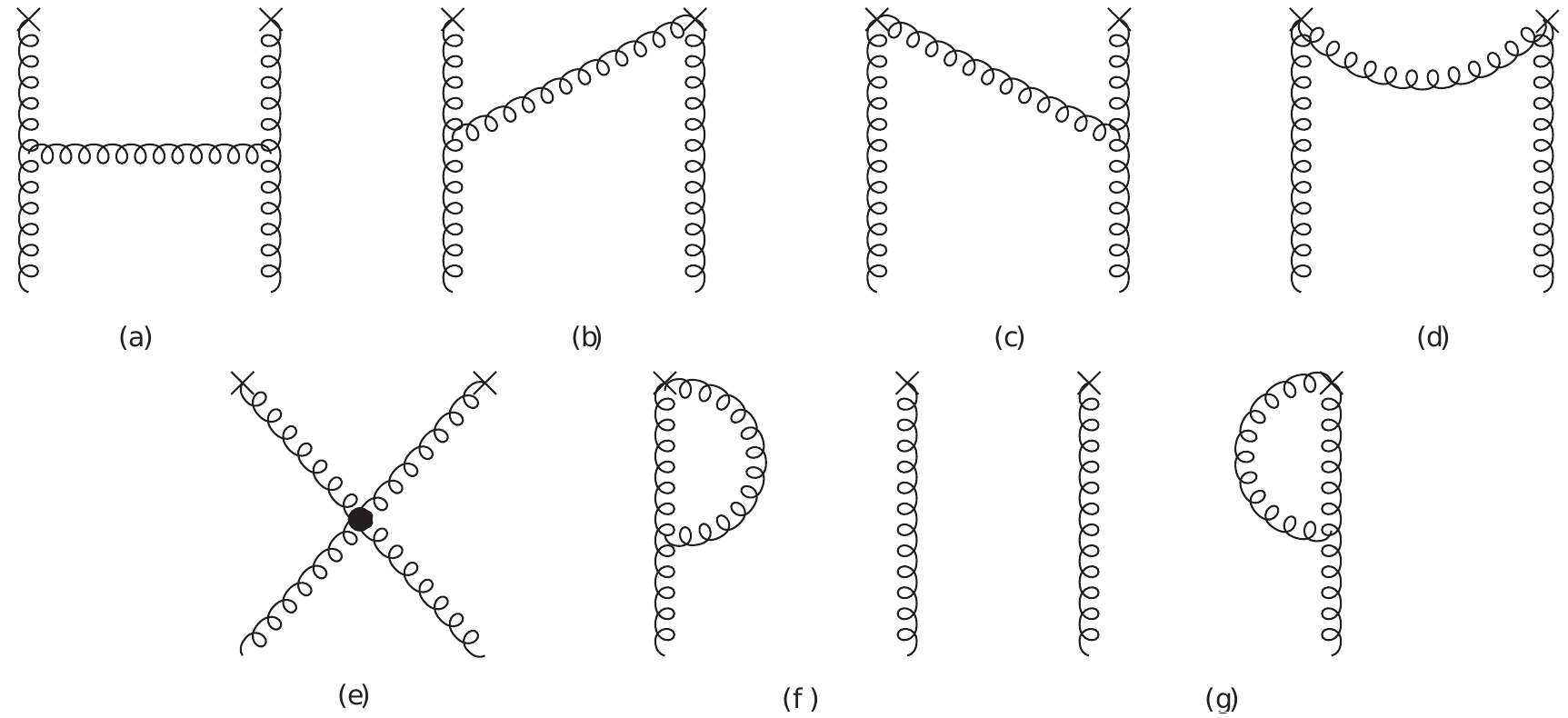}
\caption{The ``gluon in gluon" diagrams in one loop: The first set of the gluon matrix element of the equal time quasi-gluon operator defined in Eq.(\ref{O_g}). The curly lines denote the propagation of the gluons. There is a Wilson line between the crosses (``$\times$'') in each diagram contributing as an identity operator that is not shown. The gluon wave function renormalization diagram and all the crossed diagrams are not shown.
  } \label{h-I}
\end{center}
\end{figure}
Define
\begin{equation}
\label{h-I-X}
h^{B,(1)}_{gg}(\zeta,z^2)\equiv \frac{1}{\Blue{2}}\left(G_{gg}(\zeta,z^2)+G_{gg}(-\zeta,z^2) \right) ,
\end{equation}
we then list the result of each diagram \BlueX{in Fig. \ref{h-I}}: 
\begin{equation}
    \begin{split}
        G_{gg,\ref{h-I}a}(\zeta,z^2)&=\frac{\alpha_s C_A}{4\pi}\pi^{\epsilon}(-1+\epsilon)(1+\epsilon)\Gamma(-1-\epsilon)(\mu^2 z^2)^{\epsilon}\iota^{\epsilon}\int_0^1dy\ (1-y)e^{-iy\zeta}\\
        &\cdot\left[\frac{-4\epsilon-32\epsilon^2+16\epsilon^3}{\zeta^2}-\frac{2i(-\epsilon-17y\epsilon-2\epsilon^2+6y\epsilon^3)}{\zeta}+(4+2y+8y^2-8\epsilon+4y\epsilon)-i(y^2+y^3)\zeta \right]
    \end{split}
\end{equation}
with $\iota\equiv\frac{e^{\gamma_E}}{4\pi}$.
\begin{equation}
    \begin{split}
        G_{gg,\ref{h-I}b}(\zeta,z^2)=G_{gg,\ref{h-I}c}(\zeta,z^2)=\alpha_sC_A\frac{4^{-2+\epsilon}\pi^{-\frac{1}{2}+\epsilon}\Gamma(3-2\epsilon)}{\epsilon\Gamma(\frac{3}{2}-\epsilon)}(\mu^2z^2)^{\epsilon}\iota^{\epsilon}\int_0^1dy\ e^{-iy\zeta}\left[\frac{-2\epsilon-4\epsilon^2}{\zeta^2}+\frac{2i\epsilon(1+2y)}{\zeta}+(y+y^2) \right]
    \end{split}
\end{equation}
\begin{equation}
    \begin{split}
        G_{gg,\ref{h-I}d}(\zeta,z^2)=\frac{\alpha_sC_A}{\pi}\pi^{\epsilon}(1-\epsilon)\Gamma(1-\epsilon)e^{-i\zeta}\frac{1}{\zeta^2}\iota^{\epsilon}(\mu^2z^2)^{\epsilon}
    \end{split}
\end{equation}
\begin{equation}
    \begin{split}
        G_{gg,\ref{h-I}e}(\zeta,z^2)=-\alpha_sC_A\frac{4^{-1+\epsilon}\pi^{-\frac{1}{2}+\epsilon}(-1+\epsilon)(-1-4\epsilon+4\epsilon^2)\Gamma(2-2\epsilon)}{\Gamma(\frac{3}{2}-\epsilon)}\frac{1}{\zeta^2}\iota^{\epsilon}(\mu^2z^2)^{\epsilon}
    \end{split}
\end{equation}
\begin{equation}
    \begin{split}
         G_{gg,\ref{h-I}f}(\zeta,z^2)=G_{\ref{h-I}g}(\zeta,z^2)=\alpha_sC_A\frac{3\cdot2^{-3+\epsilon}e^{-i\zeta}\pi^{-1+\epsilon}(-1+\epsilon)}{\Gamma(2-\epsilon)}\left[\frac{1}{\epsilon_{UV}}-\frac{1}{\epsilon_{IR}} \right]\iota^{\epsilon}(\mu^2z^2)^{\epsilon}
    \end{split}
\end{equation}
The wave function renormalization contribution:
\begin{equation}
\label{WF}
    G_{gg,\mathrm{WF}}(\zeta,z^2)=\frac{\alpha_s}{4\pi}\left( \frac{5}{3}C_A-\frac{4}{3}T_F n_f \right)\left( \frac{1}{\epsilon_{UV}}-\frac{1}{\epsilon_{IR}} \right)e^{-i\zeta}\iota^{\epsilon}(\mu^2z^2)^{\epsilon}
\end{equation}
\begin{figure}[htp]
    \includegraphics[scale=0.6]{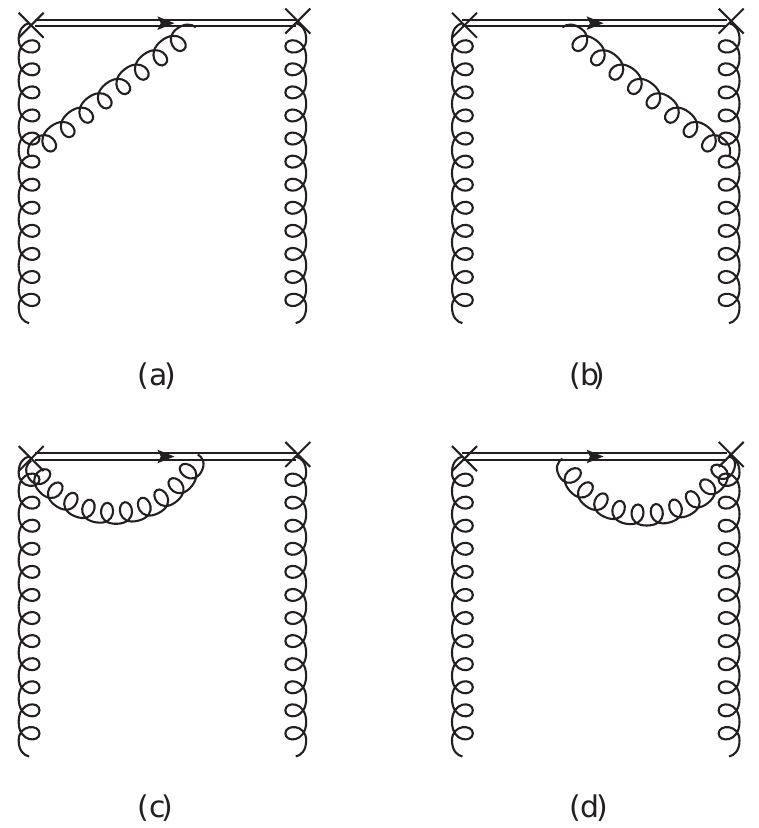}
    \caption{The ``gluon in gluon" diagrams in one loop: The second set of the gluon matrix element of the equal time quasi-gluon operator defined in Eq.(\ref{O_g}).
    The Wilson lines are denoted as the double lines. The crossed diagrams are not shown.} 
    \label{h-II}
\end{figure}
\begin{equation}
    \begin{split}
        G_{gg,\ref{h-II}a}(\zeta,z^2)&=G_{gg,\ref{h-II}b}(\zeta,z^2)\\
        &=i\frac{\alpha_sC_A}{\pi}\frac{\pi^{\epsilon}}{\zeta}\int_0^1dy\ e^{-i\zeta}\int_0^1dt\ e^{iy\zeta t}t^{-2+2\epsilon}(\mu^2z^2)^{\epsilon}\iota^{\epsilon}\\
        &+\frac{\alpha_sC_A}{2\pi}e^{\gamma_E\epsilon}(1-\epsilon)e^{-i\zeta}(i\zeta)\int_0^1dy\int_0^1dt\ e^{i\zeta t y}(\frac{t^2\mu^2z^2}{4})^{\epsilon}\Gamma(-\epsilon)\\
        &+\frac{\alpha_sC_A}{4\pi}e^{\gamma_E\epsilon}(1-\epsilon)i\int_0^1dy\ e^{iy\zeta}\frac{1}{\zeta}(-2\epsilon-i\zeta(-3+y))(\frac{\mu^2z^2}{4})^{\epsilon}\Gamma(-\epsilon)\\
        &+\frac{5\alpha_sC_A}{8\pi\Gamma(2-\epsilon)}4^{\epsilon}\pi^{\epsilon}(1-\epsilon)\iota^{\epsilon}(\mu^2z^2)^{\epsilon}\left[\frac{1}{\epsilon_{UV}}-\frac{1}{\epsilon_{IR}} \right]e^{-i\zeta}
    \end{split}
\end{equation}
\begin{equation}
    \begin{split}
        G_{gg,\ref{h-II}c}(\zeta,z^2)=G_{gg,\ref{h-II}d}(\zeta,z^2)=i\frac{\alpha_sC_A}{\pi}\frac{\pi^{\epsilon}}{\zeta}\int_0^1dy\ e^{-i\zeta}\int_0^1dt\ (-1)t^{-2+2\epsilon}(\mu^2z^2)^{\epsilon}\iota^{\epsilon}
    \end{split}
\end{equation}
Although each diagram in Fig.\ref{h-II} is divergent when $d=3$ (or $\epsilon=1/2$), their sum is free from this divergence. Only the $d=4$ divergence remains. This is also observed in Refs.\cite{Wang:2019tgg} and \cite{Balitsky:2019krf}. 


\begin{figure}[!h]
\begin{center}
\includegraphics[scale=0.6]{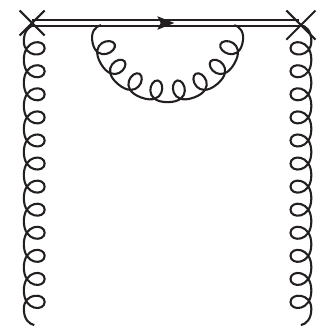}
\caption{The Wilson line self energy diagram. The crossed diagram is not shown.} \label{fig:wl2}
\end{center}
\end{figure}
For the Wilson line self energy diagram in Fig.\ref{fig:wl2}),
\begin{equation}
    \begin{split}
        G_{gg,\ref{fig:wl2}}(\zeta,z^2)&=\frac{\alpha_sC_A}{2\pi}\frac{2^{-2\epsilon}e^{-i\zeta}e^{\epsilon\gamma}(-1+\epsilon)\Gamma(1-\epsilon)}{\epsilon_{UV}(-1+2\epsilon)}(\mu^2z^2)^{\epsilon} .
    \end{split}
\end{equation}
It is linear divergent indicated by the divergence at $d=3$ or $\epsilon=1/2$ in the dimensional regularization.

\subsection{Quark in gluon}

\begin{figure}[!htp]
\includegraphics[width=0.15\textwidth]{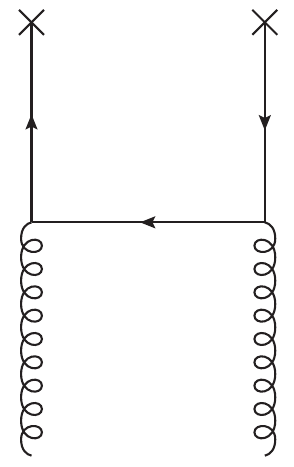}
\caption{The ``quark in gluon" diagram in one loop: the qluon matrix element of the equal time quasi-quark operator defined in Eq.(\ref{Oq}). The solid lines denote the quark propagation. The crossed diagram is not shown.}
\label{fig:gluon2quark}
\end{figure}

Define
\begin{equation}
\label{h-function-II-1}
h^{B,(1)}_{qg}(\zeta,z^2)\equiv G_{qg}(\zeta,z^2)+G_{qg}(-\zeta,z^2)  ,
\end{equation}
\BlueX{then the ``quark in gluon" diagram of Fig. \ref{fig:gluon2quark} yields}
\begin{equation}
    \begin{split}
        G_{qg}(z^2,\zeta)&=-\frac{\alpha_sT_f}{\Blue{4}\pi}\pi^{-1/2-\epsilon}2^{-2\epsilon}e^{\epsilon \gamma_E}\Gamma(\frac{3}{2}-\epsilon)\Gamma(-\epsilon)(\mu^2z^2)^{\epsilon}\int_0^1dy\ e^{-iy\zeta}\\
        &\cdot \left[-(-2\epsilon)(2-2\epsilon)(-y+y^2)+((4-2\epsilon)(-3+(3-2y)y)+4(3+(-3+y)y))+2i\zeta y(-1+y) \right]
    \end{split}
\end{equation}

\subsection{Gluon in quark}

\begin{figure}[htp]
\includegraphics[width=0.15\textwidth]{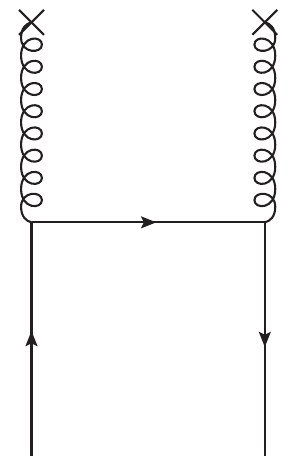}
\caption{The ``gluon in quark" diagram in one loop: The quark matrix element of the equal time quasi-gluon operator defined in Eq.(\ref{O_g}). The Wilson line between the crosses is not drawn. The crossed diagram is not shown.}
\label{fig:quark2gluon}
\end{figure}


\BlueX{Define}
\begin{equation}
\label{h-function-III}
h^{B,(1)}_{gq}(\zeta,z^2)\equiv G_{gq}(\zeta,z^2)+G_{gq}(-\zeta,z^2)  ,
\end{equation}
\BlueX{then the ``gluon in quark" diagram of Fig. \ref{fig:quark2gluon}
yields}

\begin{equation}
    \begin{split}
        G_{gq}(\zeta,z^2)&=i\frac{\alpha_sC_F}{4\pi}e^{\epsilon\gamma_E}2^{-2\epsilon}(\mu^2z^2)^{\epsilon}\Gamma(-\epsilon)\int_0^1dy\ e^{-iy\zeta}\frac{1}{\zeta}[ 4\epsilon(2+\epsilon-2\epsilon^2+y(-1+\epsilon)(1+2\epsilon))\\
        &-4i(-2\epsilon+y(2+y(-1+\epsilon)-\epsilon)(1+2\epsilon))\zeta+(-1+y)y(-4+y(2-2\epsilon))\zeta^2 ]
    \end{split}
\end{equation}

\subsection{Quark in quark}

\begin{figure}[!htp]
\centering
\includegraphics[scale=0.5]{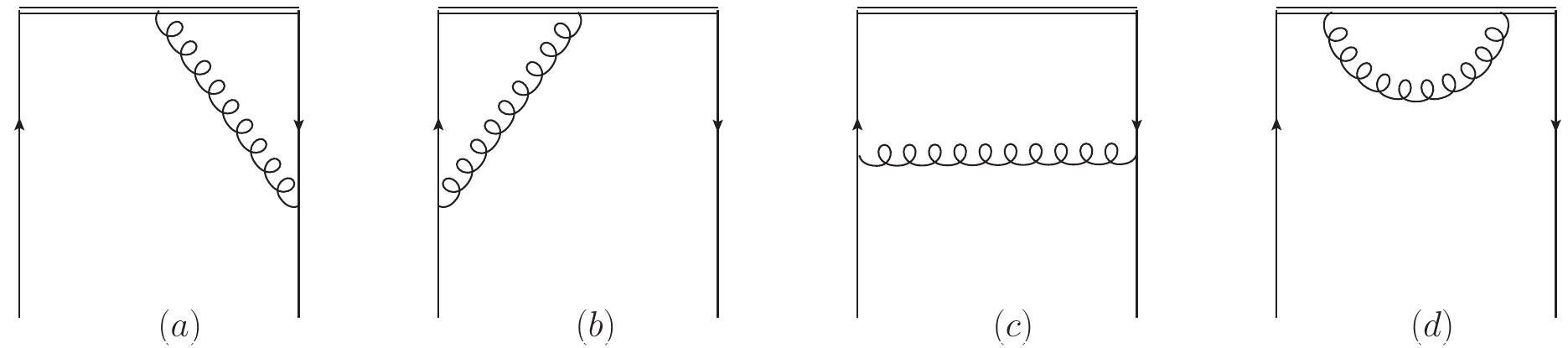}
\caption{The ``quark in quark" diagrams in one loop: The quark matrix element of the equal time quasi-quark operator defined in Eq.(\ref{Oq}). The quark wave function renormalization diagram is not shown. 
} \label{fig:quark2quark}
\end{figure}
\BlueX{Define}
\begin{equation}
\label{h-function-IV}
h^{B,(1)}_{qq}(\zeta,z^2)\equiv G_{qq}(\zeta,z^2)  ,
\end{equation}
\BlueX{then each of the ``quark in quark" diagram in Fig. \ref{fig:quark2quark} yields}

\begin{equation}
    \begin{split}               
     G_{qq,\ref{fig:quark2quark}a}(\zeta,z^2)=G_{\ref{fig:quark2quark}b}(\zeta,z^2)=&\frac{\alpha_sC_F}{2\pi}  e^{\epsilon\gamma_E} \left(i\frac{d}{d\zeta}\right) \left[4^{-\epsilon}(z^2\mu^2)^{\epsilon}(i\zeta) \int_0^1 du \int_0^1 dt (2-u) e^{-i(1-ut)\zeta} \Gamma(-\epsilon)(t^2)^{\epsilon}\right. \\
        &\left. -4^{-\epsilon}(z^2\mu^2)^{\epsilon}\int_0^1 du \ e^{-iu \zeta}\Gamma(-\epsilon) + 4^{-\epsilon}(z^2\mu^2)^{\epsilon}\left(\frac{1}{\epsilon_{\mbox{\tiny UV}}} - \frac{1}{\epsilon_{\mbox{\tiny IR}}} \right) e^{-i\zeta}\right]
    \end{split}
\end{equation}
\begin{equation}
    \begin{split}
        G_{qq,\ref{fig:quark2quark}c}(\zeta,z^2)=&\frac{\alpha_sC_F}{2\pi} e^{\epsilon\gamma_E} \left(i\frac{d}{d\zeta}\right) \int_0^1 du\ (1-\epsilon)(1-u) e^{-iu \zeta}\Gamma(-\epsilon) 4^{-\epsilon} \big(\mu |z|\big)^{2\epsilon}
    \end{split}
\end{equation}
\begin{equation}
    \begin{split}
        G_{qq,\ref{fig:quark2quark}d}(\zeta,z^2)=\frac{\alpha_sC_F}{2\pi} e^{\epsilon\gamma_E}\left(i\frac{d}{d\zeta}\right)\left(\frac{\mu|z|}{2}\right)^{2\epsilon}  \frac{\Gamma(1-\epsilon)}{\epsilon_{\mbox{\tiny UV}}} (1-2\epsilon)e^{-i\zeta}
    \end{split}
\end{equation}

\Red{Our result agrees with that of Ref.~\cite{Izubuchi:2018srq} before taking the derivative of the quark operator defined in Eq.(\ref{Oq}).}

\section{Some useful integrals} 
\label{B}

We collect some useful results of integration in this Appendix. Starting with the Fourier transform
\begin{equation}
\label{B1}
    \begin{split}
        &\int\frac{d\zeta}{2\pi}e^{ix\zeta}\frac{1}{\zeta}(\zeta^2)^{\epsilon}\Gamma(-\epsilon)4^{-\epsilon}\\
        =&\int\frac{d\zeta}{2\pi}e^{ix\zeta}\frac{|\zeta|}{\zeta}\frac{\int_{0}^{\infty}d\alpha\ \alpha^{-\frac{1}{2}-\epsilon}e^{-\alpha\zeta^2}}{\Gamma(\frac{1}{2}-\epsilon)}\Gamma(-\epsilon)4^{-\epsilon}\\
        =&4^{-\epsilon}\frac{\Gamma(-\epsilon)}{\Gamma(\frac{1}{2}-\epsilon)}\frac{ix}{2\pi}\int_0^1dt\int_0^{\infty}d\alpha\ \alpha^{-\frac{3}{2}-\epsilon}e^{-\frac{x^2}{4\alpha}(1-t^2)}\\
        =&4^{-\epsilon}\frac{\Gamma(-\epsilon)}{\Gamma(\frac{1}{2}-\epsilon)}\frac{ix}{2\pi}2^{1+2\epsilon}|x|^{-1-2\epsilon}\Gamma(\frac{1}{2}+\epsilon)\int_0^1dt\ (1-t^2)^{-\frac{1}{2}-\epsilon}\\
        =&\frac{\Gamma(\epsilon+\frac{1}{2})}{\sqrt{\pi}}i\frac{x}{|x|^{1+2\epsilon}}\frac{\Gamma(-\epsilon)}{2\Gamma(1-\epsilon)} .
    \end{split}
\end{equation}
Integrating Eq.(\ref{B1}) over $x$ yields
\begin{equation}
        \int_{-\infty}^{\infty}\frac{d\zeta}{2\pi}e^{ix\zeta}\frac{1}{\zeta^2}(\zeta^2)^{\epsilon}\Gamma(-\epsilon)4^{-\epsilon}
        =\frac{\Gamma(\epsilon+\frac{1}{2})}{\sqrt{\pi}}\frac{\Gamma(-\epsilon)}{2\Gamma(1-\epsilon)}(-1)\frac{|x|^{1-2\epsilon}}{1-2\epsilon} .
\end{equation}
Taking the $x$ derivative of Eq.(\ref{B1}) yields
\begin{equation}
    \int_{-\infty}^{\infty}\frac{d\zeta}{2\pi}e^{ix\zeta}(\zeta^2)^{\epsilon}\Gamma(-\epsilon)4^{-\epsilon}=\frac{\Gamma(\epsilon+\frac{1}{2})}{\sqrt{\pi}}\frac{1}{|x|^{1+2\epsilon}} .
\end{equation}
Taking one more $x$ derivative yields
\begin{equation}
    \int_{-\infty}^{\infty}\frac{d\zeta}{2\pi}e^{ix\zeta}\zeta(\zeta^2)^{\epsilon}\Gamma(-\epsilon)4^{-\epsilon}=(-i)\frac{\Gamma(\epsilon+\frac{1}{2})}{\sqrt{\pi}}(-1-2\epsilon)\left(\frac{1}{x|x|^{1+2\epsilon}}\right) ,
\end{equation}

\Red{Also, in wave function renormalization diagrams, we encounter}
\begin{equation}
    \mu^{2\epsilon}\int \frac{d^dk_{E}}{k_{E}^4}=\mu^{2\epsilon}\frac{2\pi^{d/2}}{\Gamma(d/2)}\int_{0}^{\Lambda}dk_E\ k_{E}^{-1-2\epsilon}+\mu^{2\epsilon}\frac{2\pi^{d/2}}{\Gamma(d/2)}\int_{\Lambda}^{\infty}dk_E\ k_{E}^{-1-2\epsilon} .
\end{equation}
Taking the scale $\Lambda=1/|z|$ to divide the IR and UV part, then we have
\begin{equation}
    \begin{split}
        \mu^{2\epsilon}\int \frac{d^dk_{E}}{k_{E}^4}&=\frac{2\pi^{d/2}}{\Gamma(d/2)}\mu^{2\epsilon}\left( \frac{1}{-2\epsilon_{IR}}z^{2\epsilon}+\frac{1}{2\epsilon_{UV}}z^{2\epsilon} \right)
=(\mu^2z^2)^{\epsilon}\frac{2\pi^{d/2}}{\Gamma(d/2)}\left( \frac{1}{2\epsilon_{UV}}-\frac{1}{2\epsilon_{IR}}\right) .
    \end{split}
\end{equation}
This was also done in Ref.~\cite{Izubuchi:2018srq}.


\section{Momentum conservation in quasi-PDFs}\label{App3}

Eq.(\ref{momcon}) shows that quasi-PDF and PDF for every parton has the same second (i.e. $\langle x \rangle$) moment. That means this tree level relation is not changed by loop corrections in the matching (similar to the quark number conservation for each quark flavor). Therefore, up to one loop matching, we should have 
\begin{equation}
    \begin{split}
        &\int x\tilde{g}(x,\mu,P^z)\ dx \\
        =& \int dx\int_{-\infty}^{\infty}\frac{d y}{|y|} \left[\xi C_{gg}(\xi,\mu,y P^z) yg(y,\mu)+ \xi C_{gq}(\xi,\mu,y P^z) y \sum_i q_i(y,\mu) \right]\\
        =&\int x g(x,\mu,P^z)\ dx\\
        &+\frac{2\alpha_sC_F}{3\pi}\int dx\int_{-\infty}^{\infty}\frac{d y}{|y|}\left\{\mathrm{ln}\left(\frac{\mu^2}{4(yP^z)^2}\right)\delta(1-\xi)-\left[\left[\frac{1}{|1-\xi|} \right]_{+(1)}^{[0,2]}+\frac{1}{|\xi-1|}\theta(-\xi)+\frac{1}{|1-\xi|}\theta(\xi-2) \right] \right\}\frac{\Red{\sum_{i}\langle x\rangle_i}}{\langle x\rangle_g}yg(y,\mu)\\
        &-\frac{2\alpha_sC_F}{3\pi}\int dx\int_{-\infty}^{\infty}\frac{d y}{|y|}\left\{\mathrm{ln}\left(\frac{\mu^2}{4(yP^z)^2}\right)\delta(1-\xi) -\left[\left[\frac{1}{|1-\xi|} \right]_{+(1)}^{[0,2]}+\frac{1}{|\xi-1|}\theta(-\xi)+\frac{1}{|1-\xi|}\theta(\xi-2) \right] \right\}y\Red{\sum_{i} q_i(y,\mu)} ,\\
    \end{split}
    \label{mcp}
\end{equation}
where $\xi=x/y$. The other plus function contributions do not contribute which can be seen easily by changing the order of the integrations. In this Appendix, we show the sum of the last two lines is zero. 

The last line of Eq.(\ref{mcp}) can be rewritten as 
\begin{equation}
\label{C2}
    \begin{split}
        &\int dx\int_{0}^{\infty}d y\left\{\mathrm{ln}\left(\frac{\mu^2}{4(yP^z)^2}\right)\delta(x-y) -\left[\frac{1}{|y|}\left[\frac{1}{|1-\frac{x}{y}|} \right]_{+(1)}^{[0,2]}+\frac{1}{|x-y|}\theta(-x)+\frac{1}{|x-y|}\theta(x-2y) \right] \right\}q_1(y,\mu)f(x)\\
        =&\int_{0}^{\infty}d y\left\{\mathrm{ln}\left(\frac{\mu^2}{4(yP^z)^2}\right)f(y) -\left[\int \frac{dx}{|y|}f(x)\left[\frac{1}{|1-\frac{x}{y}|} \right]_{+(1)}^{[0,2]}+\int_{-\infty}^{0}\frac{f(x)dx}{|x-y|}+\int_{2y}^{\infty}\frac{f(x)dx}{|x-y|}\right] \right\}q_1(y,\mu) ,
    \end{split}
\end{equation}
where the prefactor is removed. $q_1(y,\mu)\equiv y\Red{\sum_{i} q_i(y,\mu)}$, and $f(x)=1$ will be set at the end. If $y>0$, The second term of the above equation can be written as 
\begin{equation}
    \begin{split}
        &\int \frac{dx}{|y|}f(x)\left[\frac{1}{|1-\frac{x}{y}|} \right]_{+(1)}^{[0,2]}\\
        =&\int_{0}^{2} d\xi\ \frac{f(y\xi)-f(y)}{|1-\xi|}\\
        =&\int_{0}^{2y} dx\ \frac{f(x)-f(y)}{|x-y|}\\
        =&\int_{0}^{2y} dx\ \frac{f(x)}{|x-y|}-f(y)\int_{-1+y}^{1+y}\frac{dx}{|x-y|}-f(y)\ln{y^2} .
    \end{split}
\end{equation}
Therefore, Eq.(\ref{C2}) becomes
\begin{equation}
    \int_{0}^{\infty}d y\left\{\mathrm{ln}\left(\frac{\mu^2}{4(P^z)^2}\right)f(y) -\left[\int dx\ f(x)\left[\frac{1}{|x-y|} \right]_{+(y)}^{[-1+y,1+y]}+\int_{-\infty}^{-1+y}\frac{f(x)dx}{|x-y|}+\int_{1+y}^{\infty}\frac{f(x)dx}{|x-y|}\right] \right\}q_1(y,\mu)
    \label{70}
\end{equation}
This expression comes from the last line of Eq.(\ref{mcp}). In the limit of $f(x)=1$, 
its first term cancels with the counterpart in the second to the last line of Eq.(\ref{mcp}). The second term is a plus function whose integral vanishes. The last two terms have the property 
\begin{equation}
    \int_{-\infty}^{-1+y}\frac{dx}{|x-y|}+\int_{1+y}^{\infty}\frac{dx}{|x-y|}=\int_{-\infty}^{-1}\frac{dx}{|x|}+\int_{1}^{\infty}\frac{dx}{|x|} ,
    \label{C5}
\end{equation}
which again cancels the counterparts of the second to the last line of Eq.(\ref{mcp}). Therefore, we have shown the last two lines of Eq.(\ref{mcp}) sums up to zero for $y>0$. Similarly, one can show that the cancelation also happens for $y<0$.

\end{widetext}

\bibliography{main_V4.bbl}
\end{document}